\documentclass{aa}
\usepackage{epsfig}
\usepackage{epstopdf}
\usepackage{amsmath}
\usepackage{amssymb}
\usepackage{subfigure}
\usepackage{float}
\usepackage[usenames]{color}
\usepackage{color,epsfig,amsmath,amssymb,fancyhdr}   
\usepackage{ulem}  
\usepackage{url,hyperref}
\usepackage{indentfirst}

\usepackage{footmisc}
\usepackage{tabularx}
\usepackage{natbib}
\usepackage[switch, mathlines]{lineno} % not working with revtex4 !

\makeatletter
\renewcommand*{\@fnsymbol}[1]{\ifcase#1\or*\or$\dagger$\or$\ddagger$\or**\or$\dagger\dagger$\or$\ddagger\ddagger$\fi}
\makeatother

\usepackage[usenames]{color}
\definecolor{Cedric}{rgb}{1.,0.,0.}

\graphicspath{ {./Figures/} }

\begin{document}

\title{Modeling spectral lags in active galactic nucleus flares in the context of Lorentz invariance violation searches}
\titlerunning{Spectral lags in AGN flares}

\author{C. Perennes\inst{\ref{LUTH},\ref{LPNHE}, \ref{Padova}}, H. Sol\inst{\ref{LUTH}} and J. Bolmont\inst{\ref{LPNHE}}}

\institute{
        LUTH, Observatoire de Paris, PSL Research University, CNRS, Universit\'e Paris Diderot, 5 Place Jules Janssen, F-92190 Meudon, France \label{LUTH}
   \and
        Sorbonne Universit\'e, Universit\'e Paris Diderot, Sorbonne Paris Cit\'e, CNRS/IN2P3, Laboratoire de Physique Nucl\'eaire et de Hautes Energies, LPNHE, 4 Place Jussieu, F-75252 Paris, France \label{LPNHE}
        \and
        Presently at Universit\`a di Padova and INFN, I-35131 Padova, Italy \label{Padova} 
}

\date{Received August 1, 2019; accepted November 12, 2019.}

% \abstract{}{}{}{}{} 
% 5 {} token are mandatory
 
 \abstract
 % context heading (optional)
 % {} leave it empty if necessary
{High-energy photons emitted by flaring active galactic nuclei (AGNs) have been used for many years to constrain modified dispersion relations in vacuum encountered in the context of quantum gravity phenomenology. In such studies, done in the GeV-TeV range, energy-dependent delays (spectral lags) are searched for, usually neglecting any source-intrinsic time delay.}
% aims heading (mandatory)
{With the aim being to distinguish lorentz invariance violation (LIV) effects from lags generated at the sources themselves, a detailed investigation into intrinsic spectral lags in flaring AGNs above 100~GeV is presented in the frame of synchrotron-self-Compton (SSC) scenarios for their very-high-energy (VHE) emission.}
% methods heading (mandatory)
{A simple model of VHE flares in blazars is proposed, allowing to explore the influence of the main physical parameters describing the emitting zones on intrinsic delays.}
% results heading (mandatory)
{For typical conditions expected in TeV blazars, significant intrinsic lags are obtained, which can dominate over LIV effects, especially at low redshifts, and should therefore be carefully disentangled from any extrinsic lags. Moreover, two main regimes are identified with characteristic spectral lags, corresponding to long-lasting and fast particle acceleration.}
% conclusions heading (optional), leave it empty if necessary
{Such intrinsic spectral lags should be detected with new-generation instruments at VHE such as the Cherenkov Telescope Array which begins operation in a few years. This will provide original constraints on AGN flare models and open a new era for LIV searches in the photon sector.}

%\keywords{gamma-rays, blazar, flares, time lags, modeling, Lorentz Invariance Violation, Quantum Gravity}

\maketitle

%\linenumbers

\section{Introduction}
\label{sec:intro}

Energy-dependent time-lags in signals arriving from remote cosmic gamma-ray emitters are of particular interest both for understanding the physics of astrophysical sources and for investigating possible new phenomena impacting on photon propagation. Lorentz invariance violation (LIV) is an example of such a phenomenon. It appears as a striking outcome of some Quantum Gravity (QG) models in the form of a modified dispersion relation for photons in vacuum and is one of the most explored ways in QG phenomenology \citep{Amelino2013}. It is also included as a starting hypothesis in the standard model extension (SME), an effective field theory built from the Lagrangian of the standard model of particle physics including terms for LIV and charge parity time reversal (CPT) violation \citep{Kostelecky2008, Kostelecky2009}.

Although the present paper focuses on spectral lagsi nduced by a LIV effect in the photon sector, it is necessary to stress that these are not the only possible phenomenon arising from LIV that could be measured with astrophysical sources \citep[see][for a review]{Mattingly2005}. A modified dispersion relation for photons in vacuum can indeed be interpreted as the photon taking a nonzero effective mass. In that case, normally forbidden processes such as photon decay ($\gamma \to e^+ e^-$) and Cherenkov radiation ($e^- \to \gamma e^-$) would be allowed in vacuum, and cross-section for the $\gamma_\mathrm{HE} + \gamma_\mathrm{EBL}$ absorption process of high-energy photons on the extragalactic background light (EBL) would be modified. The latter effect would result in the Universe being more transparent than expected with the standard EBL absorption with no LIV \citep{bit:2015,tav:2016,abd:2019}. Depending on the QG model considered, it is also possible to obtain vacuum birefringence in addition to energy-dependent speed of light: photons could have different speeds depending on their polarization \citep{got:2014,wei:2019}. Several other types of effects are also actively analyzed in the electron sector \citep{alt:2005,alt:2006} and in the gravitational sector \citep{Leponcin2016,Bourgoin2017,Kostelecky2017}. Such LIV effects might further modify the launching mechanism of active galactic nucleus (AGN) jets and the radiative processes involved in the generation of blazar flares, but are neglected in the present work.

The first sources proposed to look for LIV spectral lags were gamma-ray bursts (GRBs) because they could be observed at large redshifts and in great numbers by satellites in the soft gamma-ray range \citep{ame:1998}, but flaring AGNs were used almost from the same time \citep{Biller1999}. As in GRBs, AGNs appear to be strong gamma-ray emitters and their active states are observed by detectors with a high effective area such as ground-based Imaging Atmospheric Cherenkov Telescopes (IACTs), with large enough sample sizes to measure fast variability. In addition, AGNs can be monitored regularly and since an AGN flare lasts longer than a GRB,  the probability to catch a flare signal under alert is higher. In this regard, blazars emitting in the tera-electronvolt (TeV) range are especially interesting since they are the most variable population of gamma-ray loud AGNs. Their observation with the first generation of  IACT, such as Whipple, H.E.S.S., MAGIC, and VERITAS, has already provided stringent limits on spectral time-lags and LIV parameters \citep{Albert2008, abr:2011, abr:2015, zit:2013}. 

The performance aimed for CTA, the Cherenkov Telescope Array \citep{ach:2017}, will potentially allow for increasingly significant lag measurements thanks to a  larger energy range \mbox{(20~GeV -- 300~TeV)}, a higher sensitivity ($\times$10), and a better temporal resolution with respect to the present generation of IACT, and observation strategies designed to optimize the number of transient or variable object detections. The measured lags, if  significant, will have to be interpreted as propagation delays, as effects intrinsic to the sources, or as a superposition of both. 

The present work is a first attempt to gain knowledge on source-intrinsic spectral lags of flaring AGNs at high and very high energies and on short timescales relevant for LIV searches, using leptonic AGN flare modeling. In this study, only the LIV effect on the propagation of photons in vacuum and the subsequent time-delays engendered are considered to keep a relatively simple AGN flare model. Section 2 presents the general context of the search for LIV signatures from the analysis of blazar gamma-ray flares. We briefly present and apply a standard Synchrotron-Self-Compton (SSC) scenario for such flares in section 3, focusing on purpose on the dominant and unavoidable mechanisms needed to generate the burst. We explore the intrinsic SSC time delays which are induced in the gamma-ray range in section 4, and further characterize them in section 5. Section 6 focuses on the VHE domain (E > 100 GeV) to compare potential intrinsic and LIV time-lags,  and further astrophysical issues are discussed. Conclusions and perspectives are presented in section 7.

\section{Search for Lorentz invariance violation from spectral lags in blazars}
\label{sec:liv}
\subsection{Spectral lags and Lorentz invariance violation}

Focusing on spectral lags in the context of QG phenomenology, and neglecting birefringence effects, the modified dispersion relation is usually expressed as

\begin{equation}
\label{eq:disprel1}
E^2 \simeq p^2 c^2\times\left[1 \pm \ \left(\frac{E}{E_{QG}}\right)^n\right],
\end{equation}
where $c$ is the low-energy limit of the speed of light, $E_{QG}$ is the energy to be measured or constrained at which LIV effects should become non-negligible and the sign $\pm$ translates into the possibility to have an increasing (superluminal) or decreasing (subluminal) speed when photon energy increases. The value of $E_{QG}$ is usually expected to be of the order of the Planck scale \mbox{$E_P \sim 10^{19}$ GeV}. Present day IACT experiments probe the linear effect (n=1) while the quadratic effect (n=2) is still far from reach. However, other orders are investigated at VHE with other experiments \cite[see][and references therein]{kos:2011}. Moreover, for odd values of $n$, there is a correspondence between the photon helicity and the subluminal or superluminal case. In this first work, we neglect this helicity effect and assume only either the subluminal or superluminal case for odd $n$ values.

The use of variable or transient and distant astrophysical sources for LIV searches was first proposed in the late~90s \citep{ame:1998}. The modified dispersion relation of Equation~\ref{eq:disprel1} naturally leads to an energy-dependent group velocity of light. Two photons of different energies ($E_h$ and $E_l$, with $E_h > E_l$), assumed to have been emitted at the same time from the same place by a source at redshift $z$ would be detected with a spectral lag

\begin{equation}
\label{eq:timez5}
\Delta t_{\mathrm{LIV},n} \simeq \pm\,\frac{n+1}{2}\,\frac{E_h^n - E_l^n}{E_{QG}^n}\ \kappa_n(z),
\end{equation}
where the distance parameter 

 \begin{equation}
\kappa_n(z) = \int_0^z \frac{(1+z')^n}{H(z')} dz'
 \end{equation}
is an increasing function of redshift taking into account Universe expansion during photon propagation \citep{Jacob2008}. Here,$H(z)$ is the Hubble parameter. It is necessary to point out that this expression, although used in all LIV searches performed so far, was obtained under the implicit assumption that translations are not affected by Planck scale effects. Other expressions have been proposed, for example in the deformed special relativity (DSR) approach \citep[see e.g.,][]{ros:2015}. From Equation~\ref{eq:timez5}, it is common to express the “time-lag over energy difference'' parameter $\tau_n$ such that

\begin{equation}
\tau_n \equiv \frac{\Delta t_{\mathrm{LIV},n}}{E_h^n - E_l^n}.
\end{equation}
This parameter is constrained and limits on $E_{QG}$ are derived from astrophysical source observations.

The lags induced by a LIV effect are expected to be small. Maximizing them requires observation of sources preferably with high redshifts and hard spectra on a wide energy range so that $\kappa_n$ and $E_h^n - E_l^n$ are maximized. Fast variability is also required in order to measure the lags. The high-energy gamma-ray domain is therefore  particularly suitable for such studies. 

The expression of Equation~\ref{eq:timez5} was obtained assuming that high- and low-energy photons are emitted at the same time from the same place, that is, neglecting any source-intrinsic delays possibly resulting from emission mechanisms and source extent. In principle, the measured time-lag $\Delta t_m$ should rather be expressed as the sum of delays with different origins:

\begin{equation}
\label{eq:totdelay}
\Delta t_m = \Delta t_{\mathrm{LIV}} +(1+z)\Delta t_s + \sum_j \Delta t_j,
\end{equation}
where $\Delta t_s$ is the delay due to the emission processes at the source located at redshift $z$ and $\Delta t_j$ accounts for various additional effects which could affect  $\Delta t_m$. These terms include {for example} the dispersion by free electrons along the line of sight, mostly important in the radio range, potential lags due to special relativistic effects in the case where photons have a nonzero rest mass, or lags caused by the gravitational potential integrated from the source to the Earth \citep[see][and references therein]{gao:2015, wei:2016}. These extra terms $\Delta t_j$ are neglected in the present work.

\subsection{Intrinsic time-lags from a blazar flare}

        In the gamma-ray domain, only one flare from Mrk~501 has shown an indication of a $4\pm1$~min time-lag between energy bands below 250~GeV and above 1.2~TeV \citep{Albert2007}. Barely significant spectral lags $\tau_1 = (0.030\pm0.012)$~ s GeV$^{-1}$ and $\tau_2 = (3.71\pm2.57)\times 10^{-6}$~s GeV$^{-2}$ were later reported from the same data set \citep{Albert2008}. This flare, recorded on July 9 2005 by the MAGIC Cherenkov telescopes, suggests that intrinsic delays can exist in AGN flares, while the fact that it is the only one detected implies that intrinsic effects are certainly different for each AGN and perhaps from one flare to another, even in the same source. 

        Apart from the previous example, no significant spectral lag has ever been measured at GeV and TeV energies from AGNs, and source intrinsic effects have been ignored when constraining the LIV energy scale. Neglecting intrinsic effects could be partially justified when the energy range considered in the analysis is restricted enough to ensure that observed photons can be considered as all emitted together at once in the cosmic source. However, using stringent energy selections has a drawback since it drastically decreases statistics.

        Delays can indeed easily appear during the emission of photons from cosmic sources. Such intrinsic spectral time-lags  have already been unambiguously detected in some GRBs, and have mostly been reproduced by models considering light-travel-time effects from extended and inhomogeneous emitting zones as expected in standard GRB scenarios \citep[and references therein]{dai:2017}. In the case of blazar flares, intrinsic effects do not appear to be as important and have been poorly detected up to now. Nevertheless, they should be detected, either soon with current instruments during some extraordinary flare fully monitored over a large spectral gamma-ray range, or in the coming years with a new generation of instruments providing high-quality light curves and dynamical spectra at VHE. Therefore, in the context of LIV signature searches, intrinsic effects need to be further investigated, at least to help in constraining the QG energy scale in case of future detections of significant time delays in AGN flares \citep{per:2017}. 

        Several time-dependent emission models of nonthermal emission in AGNs have been proposed in the literature, such as for example by \citet{bla:1979}, \citet{mar:1985}, \citet{cel:1991}, \citet{bot:1997}, \citet{kat:2003}, \citet{jos:2011},\citet{lew:2016}. As a matter of fact, such scenarios usually induce possible spectral lags but such intrinsic lags have not been purposefully analyzed in the context of TeV blazar flares. We distinguish here two different types of time delays, the “macroscopic'' and the “microscopic'' ones, depending on their values and origins. 
        
        Macroscopic delays correspond to a variety of long lags induced by the global structure and properties of extended nonhomogeneous emitting zones. Different parts of the source emit in different spectral ranges, therefore inducing possibly important time-lags depending on the specific geometry and kinetics of the radiating plasma. This type of scenario was proposed for instance by \citet{sok:2004}, while triggering an AGN flare by the collision of a stationary  shock wave with a relativistic shock wave in the jet, later forming reverse and forward shocks which both accelerate particles. The complex jet geometry as well as acceleration and emission pattern which then develop can result in long time-lags. This model successfully reproduced a flare observed in 3C~273, in which X-rays were delayed by about one day with respect to their infrared counterparts \citep{sok:2004}. Another example of "macroscopic delay" has been obtained by \citet{bed:2008} who explained the time-lag of about four minutes measured in the 2005 Mrk~501 flare by considering  a continuous increase of the global Lorentz factor of the emitting zone propagating along the jet. In such a model of an accelerating blob, with an increase of the Doppler boosting effect,  lower-apparent-energy photons are, on average, emitted in a different location from higher-apparent-energy photons, creating the observed spectral time delay.
        
        In the present study we focus on microscopic delays, which are barely discussed at TeV energies although they are easily produced in standard one-zone SSC models and naturally come from the temporal evolution of the distribution of emitting particles. Such microscopic delays have already been analyzed by \citet{lew:2016} in the context of \mbox{X-ray} variability studies in blazar jets. Focusing on synchrotron emission, these latter authors produced flare light curves and their associated X-ray time-lags. Applying them to data from Mrk~421, they were able to characterize some acceleration parameters in that source. However, \citet{lew:2016} entirely neglect the inverse-Compton losses which are mandatory to describe gamma-ray time delays. In the following section, we consider a minimal time-dependent leptonic model to generate blazar flares and  study short  intrinsic time delays at \mbox{gamma-ray} energies, above 1~MeV. 

\section{Time-dependent model}
\label{sec:model}
        \subsection{Description of the electron distribution}
The time evolution of a flare is deduced from the description of the electron population $N_\mathrm{e}(\gamma,t)$ at the source in a single homogeneous emitting zone, with \mbox{$\gamma = E/(m_\mathrm{e}c^{2})$} being the Lorentz factor of electrons. We adopt the standard SSC scenario presented in \citet{kat:2001,kat:2003} but simplify the geometry to focus on $\gamma$-ray flares arising from a small spherical blob of plasma. Further developments of the simulation codes were done to adapt them to a detailed analysis of spectral lags. The evolution of the electron density is expressed as:

\begin{equation}        
\label{eq:equadiff}
\frac{\partial N_e^{*}(t,\gamma)}{\partial t} = \frac{\partial}{\partial \gamma} \left\{ \left[\gamma^2 C_{\mathrm{cool}}(t) - \gamma C_{\mathrm{\mathrm{acc}}}(t)\right]\ N_\mathrm{e}^{*}(t,\gamma) \right\},
\end{equation}
where $C_{\mathrm{cool}}$ accounts for radiative energy losses and $C_{\mathrm{acc}}$ for acceleration of electrons or energy gain (neglecting here adiabatic losses). This differential equation admits an analytical solution with an initial electron spectrum $N_\mathrm{e}(0,\gamma)$ defined as a power law function with a cut-off: 

\begin{equation}
N_\mathrm{e}(0,\gamma) = K_0 \gamma^{-n}\left[1 - \left(\frac{\gamma}{\gamma_{c,0}}\right)^{n+2}\right],
\label{eq:N0}
\end{equation}
where $K_0$ is the electron density at $\gamma = 1$, $n$ the electron spectrum index, and $\gamma_{c, 0}$ the cut-off energy. 

The coefficient $C_{\mathrm{cool}}$ takes into account the SSC energy losses and can be expressed as follows: 

\begin{equation}
        C_{\mathrm{cool}}(t) = \frac{4\sigma_T\left(U_\mathrm{B}(t) + U_r(t)\right)}{3m_\mathrm{e}c},
        \label{eq:Ccool}
\end{equation}
where the first term corresponds to synchrotron energy losses, with $U_\mathrm{B}(t) = B(t)^2/8\pi$ being the magnetic field energy density, and the second term is related to the inverse-Compton (IC) energy losses on the synchrotron photon field, corresponding to the synchrotron energy density $U_r(t)$. It is parameterized as:

\begin{equation}
U_r(t) = U_\mathrm{B}(t)/\eta.
\end{equation} 
This choice allows us to simplify the resolution of Equation~\ref{eq:equadiff}, which then admits an analytical solution, and to decrease the computation time significantly. The $\eta$ parameter represents the relative importance between the synchrotron and IC radiative cooling. It has to be constant over time or to be large enough in order for the analytic solution to be valid. This limits the  parameter space that can be explored to the one corresponding to synchrotron-dominated sources like high-frequency peaked BL Lac (HBL) objects.

We further assume a characteristic time $t_0$ for the temporal evolution of the main flare parameters. This time is typically related to the speed of sound in a blob of relativistic plasma characterized by a radius $R_0$ as $t_0 = R_0/(c\sqrt{3})$, $c$ being the speed of light in vacuum. The evolution of the magnetic field strength is then described as:

\begin{equation}
        B(t) = B_0 \left(\frac{t_0}{t}\right)^{m_\mathrm{b}}, 
        \label{eq:Bt}
\end{equation}
with $B_0$ being the initial value of the magnetic field at $t_0$, and $m_\mathrm{b}$ its temporal evolution index.

Finally, $C_{\mathrm{acc}}$ corresponds to the acceleration processes and allows us to initiate the flare starting from low states of the electron spectrum distribution. It is expressed as:

\begin{equation}
        C_{\mathrm{acc}}(t) = A_0 \left( \frac{t_0}{t} \right)^{m_a}
        \label{eq:Cacc}
,\end{equation}  
where $A_0$ is the initial acceleration amplitude, and $m_a$ the acceleration temporal index.

        \subsection{Spectral energy distribution}
        \label{sec:sed}
\begin{figure}[ht]
        \begin{center}
                \includegraphics[width=.98\linewidth]{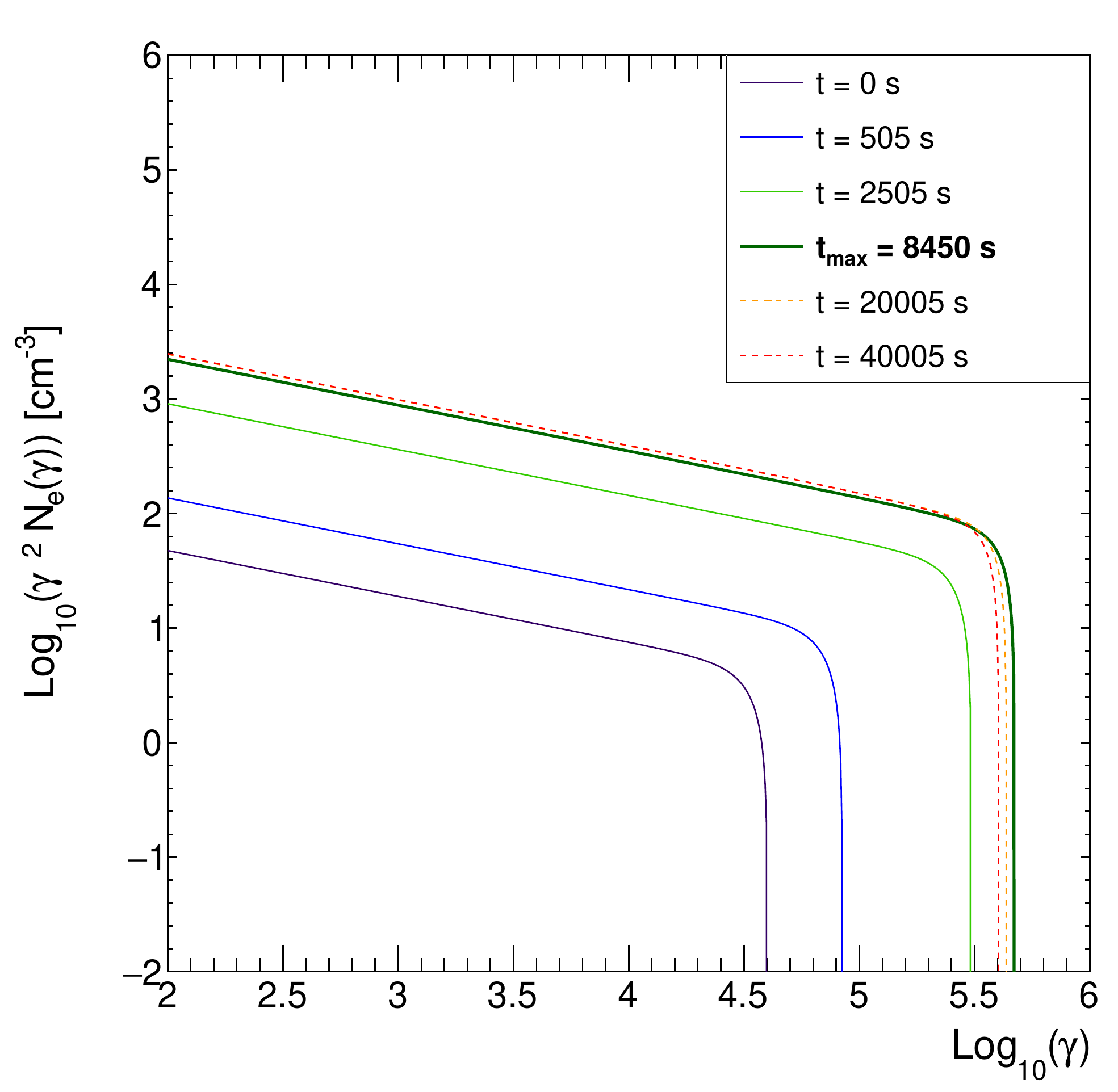}
                \includegraphics[width=.98\linewidth]{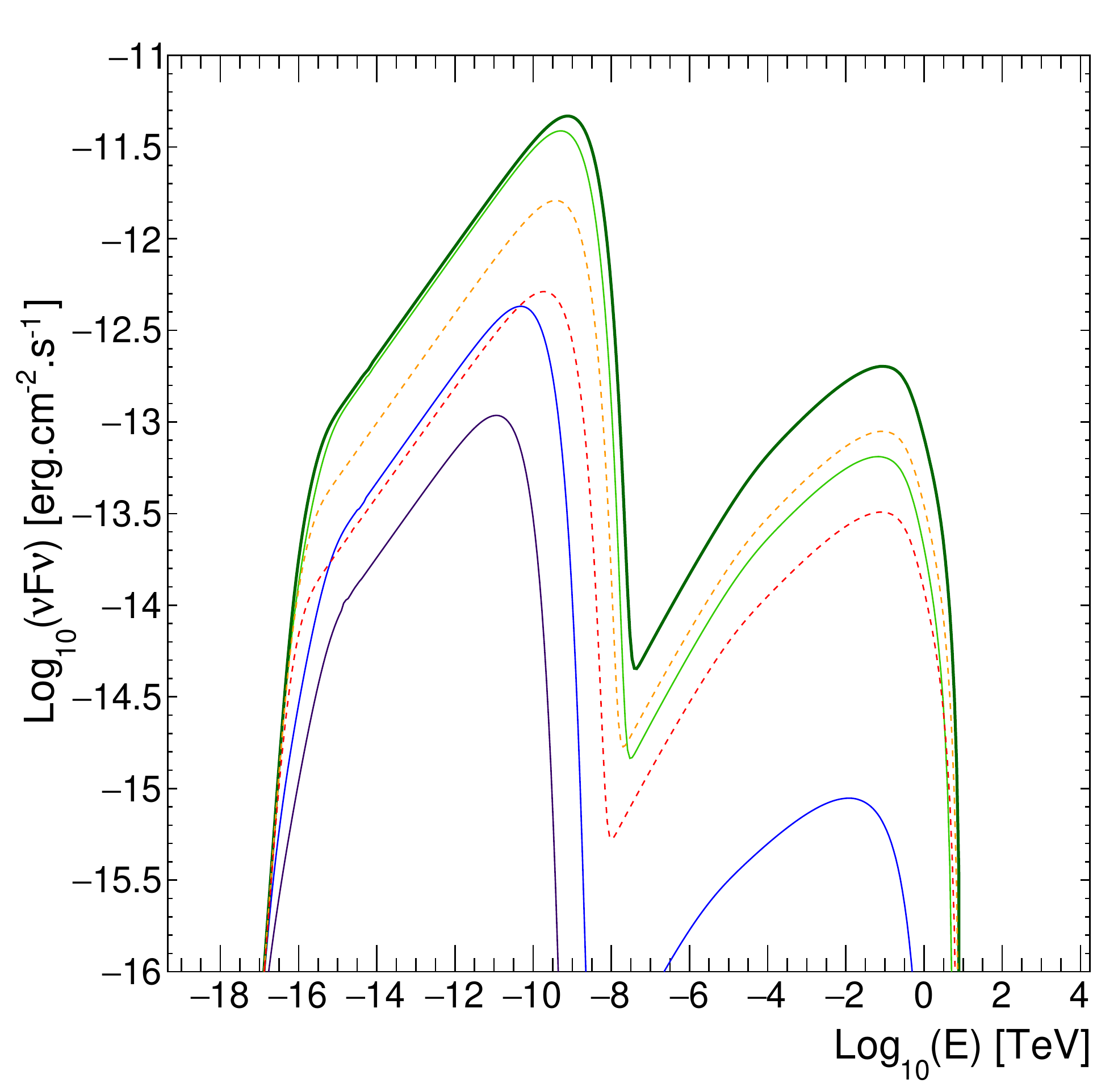}
        \end{center}
                \caption{Electron spectrum (top) and SED (bottom) evolution for the reference set of parameters given in Table~\ref{tab:par_std}, illustrating a typical TeV blazar flare. The two plots share the same color code. The full lines correspond to the rising phase of the flare and the dashed lines to its decay. The time $t_{max}$ is defined as the time when the highest energy value $\gamma_{max} = \mathrm{max}\left(\gamma_{c}(t)\right)$ is reached in the electron spectrum. The corresponding electron spectrum and SED are plotted in thick lines. The electron spectrum and SED at the maximum luminosity of the flare are not displayed because they almost coincide with the ones at $t_{max}$ for this reference set of parameters.}
                \label{fig:sed}
\end{figure}

   Spectral energy distributions (SEDs) are generated assuming a SSC emission model. To compute the SSC radiation, we use standard packages described in \citet{kat:2001} which provide the synchrotron and IC emission from a given electron spectrum, taking into account synchrotron self absorption of the emitting zone. Absorption from the extragalactic background light (EBL) is included following  the model of  \citet{fra:2008}. The evolution of the electron spectrum and of the SED  is shown in Figure~\ref{fig:sed} for the set of parameters given in Table~\ref{tab:par_std}. Such physical parameters correspond to standard values as previously deduced  with the same SSC packages and others for the archetypal TeV sources  Mrk~501, Mrk~421, and PKS 2155-304 \citep{kat:2001,kat:2003,abr:2012} and can be considered as typical values expected in blazar flares. This explains the small redshift value adopted here, leading to a low EBL absorption.

\renewcommand{\arraystretch}{1.3}
\begin{table}[ht]
        \begin{center}
                \begin{tabular}{l l l}
                        \hline
                        SSC Parameter           & Value                                         & Unit            \\ \hline       
                        $\delta$                        & 40                                             &                       \\
                        $B_0    $                               & 65                                            & mG              \\
                        $R_0$                           & $5 \times10^{15}$             & cm                      \\
                        $K_0$                           & 300                                   & cm$^{-3}$       \\
                        $\gamma_{cut}$          & $4 \times 10^{4}$             &                         \\      
                        n                                       & 2.4                                   &                         \\ 
                        z                                       & 0.03                                  &                         \\ \hline
                        Evolution parameter     & Value                                         & Unit            \\ \hline
                        $A_0$                           & $4.5 \times 10^{-5}$         & s$^{-1}$      \\
                        $m_a$                           & 5.6                                   &                         \\
                        $m_\mathrm{b}   $                               & 1                                               &                       \\ \hline          
                \end{tabular}
                \vspace{.2cm}
        \caption{Standard parameters expected in typical blazar TeV flares, considered as the first reference set throughout this paper.}
        \label{tab:par_std}
        \end{center}
\end{table}
\renewcommand{\arraystretch}{1} 

\subsection{Light curve determination}

\begin{figure*}[ht]
        \includegraphics[width=.33\linewidth]{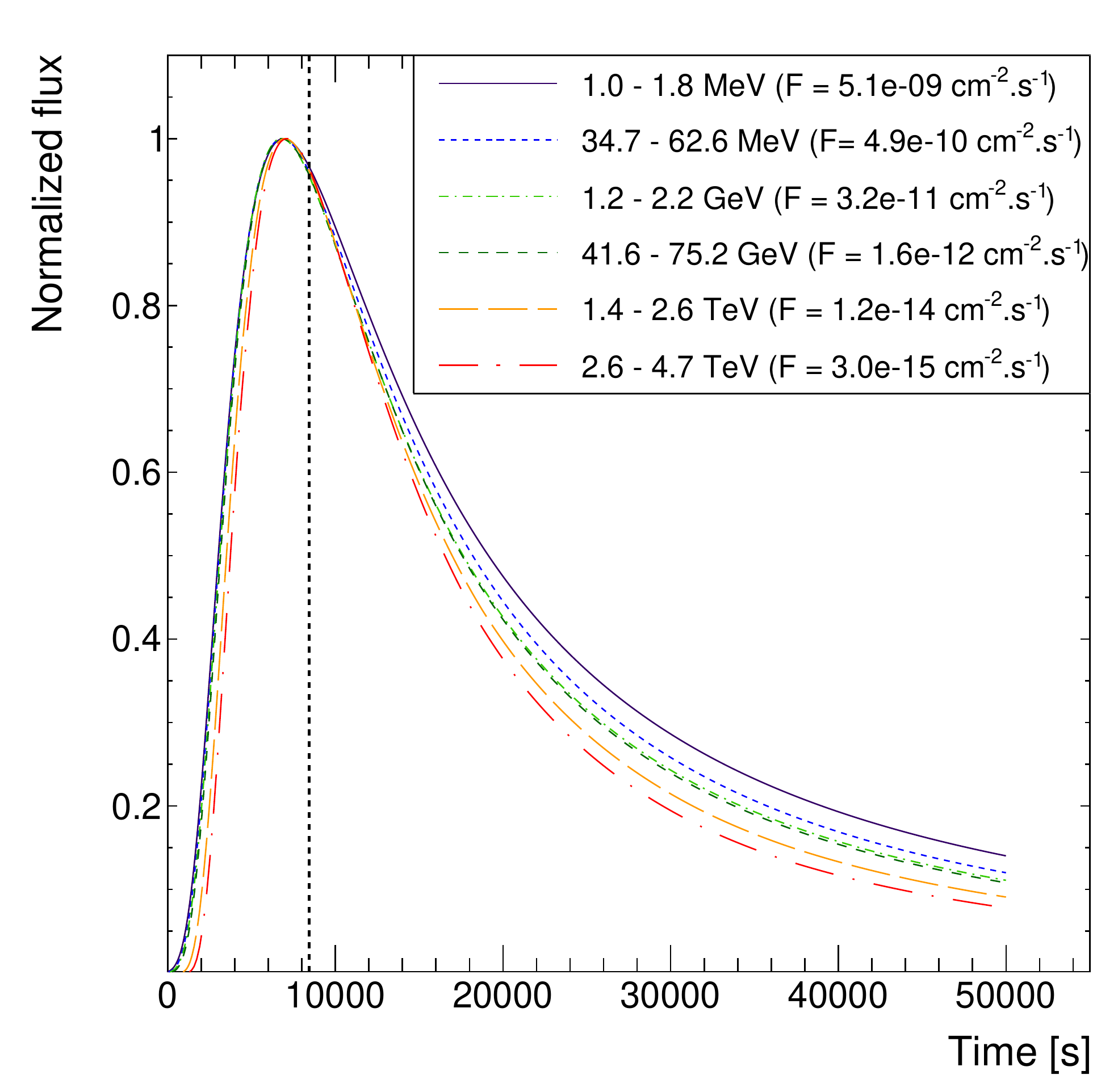}\label{fig:lc1}
        \includegraphics[width=.33\linewidth]{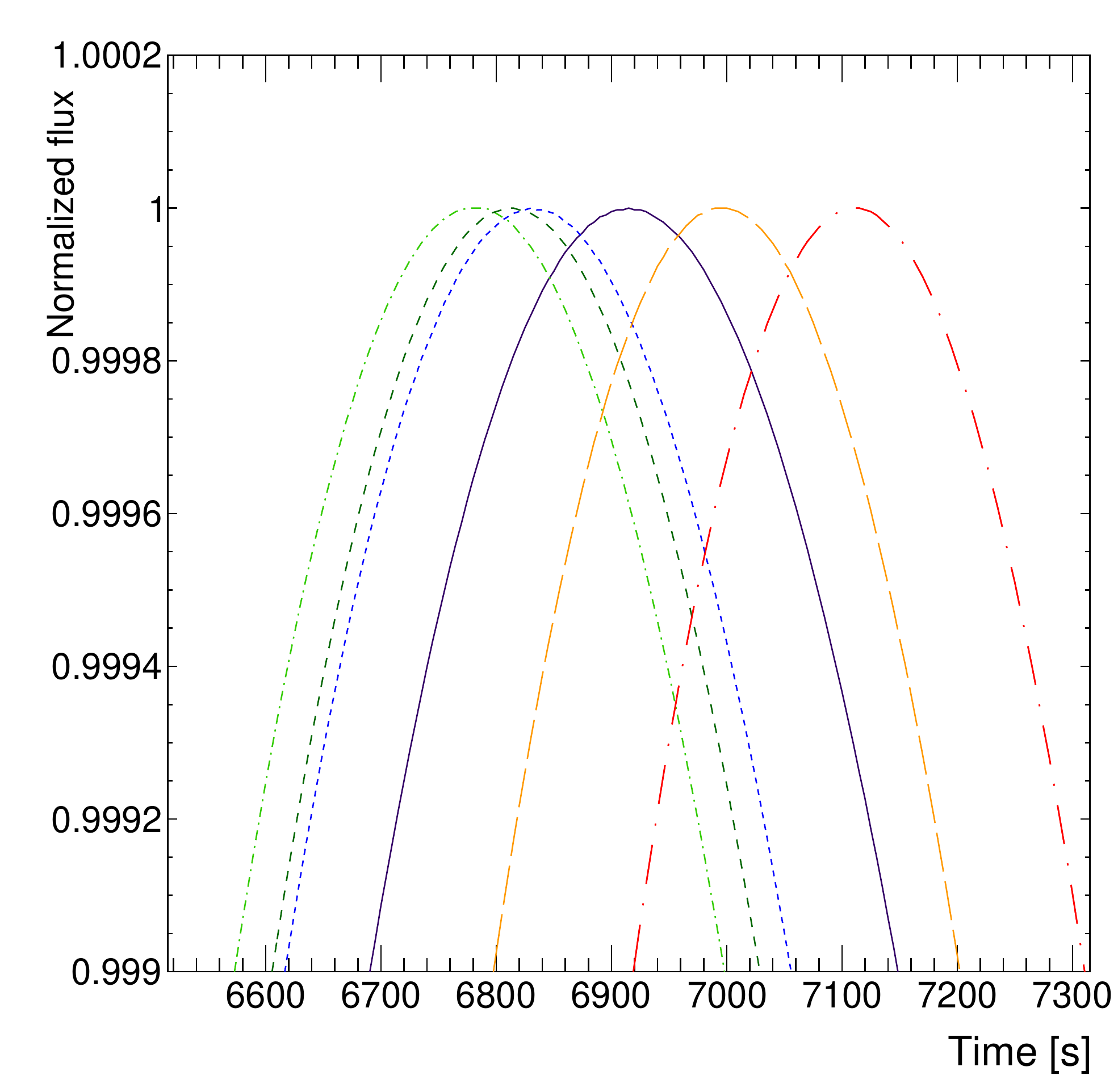}\label{fig:lc1z} 
        \includegraphics[width=.33\linewidth]{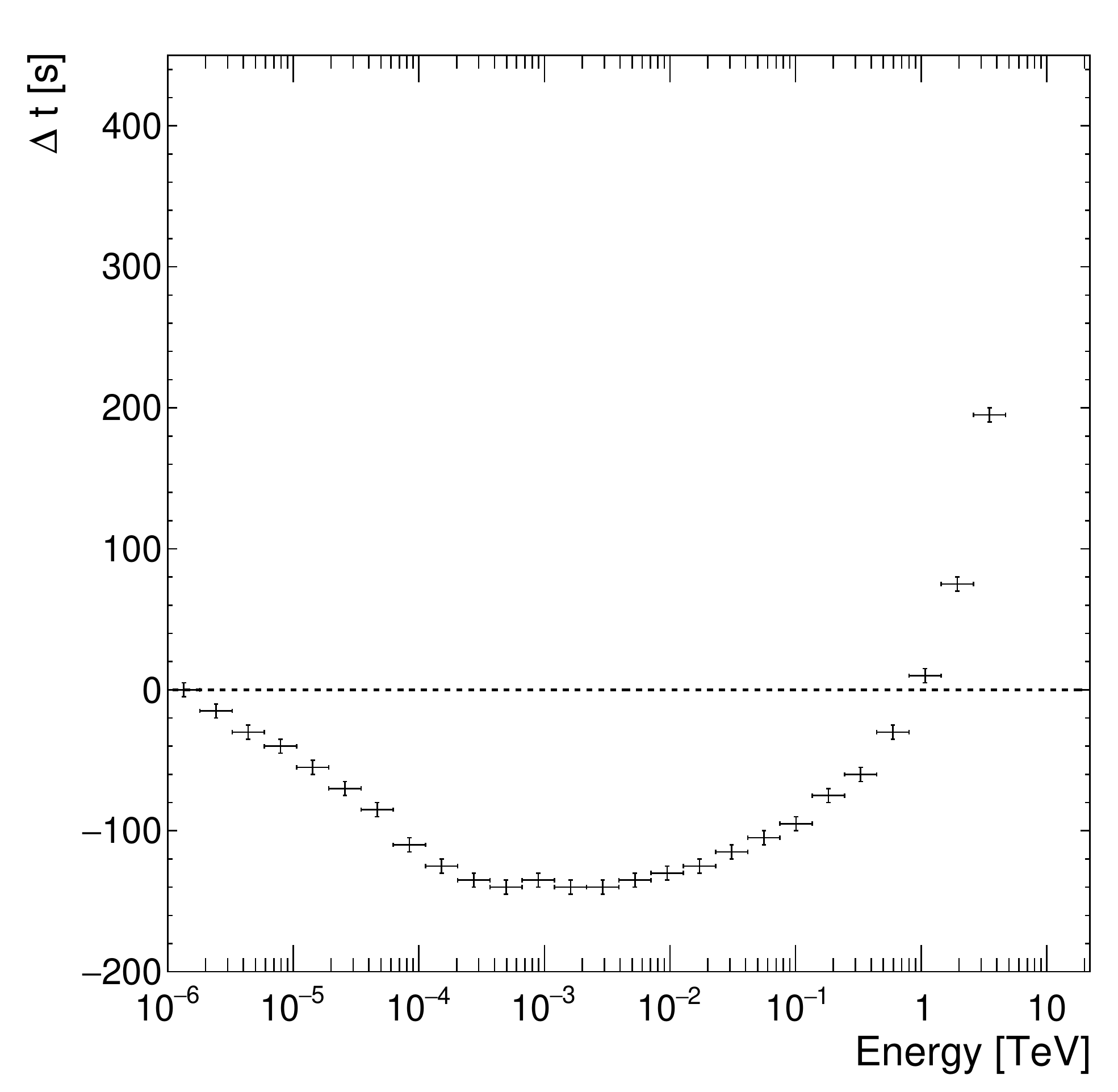}\label{fig:lag1}
\caption{Normalized light curves (left), zoom close to the light-curve maxima (center), and time-delays between each light curve and the one in the range 1-1.8 MeV (right) obtained for the first reference set of parameters. The vertical dashed line on the light curves (left) corresponds to the time when electrons reach their maximum energy $\gamma_{max}$ and shows the moment when energy losses dominate over the acceleration at the energy $\gamma_{max}$.}
\label{fig:cas1}
\end{figure*}

\begin{figure*}[ht]
        \includegraphics[width=.33\linewidth]{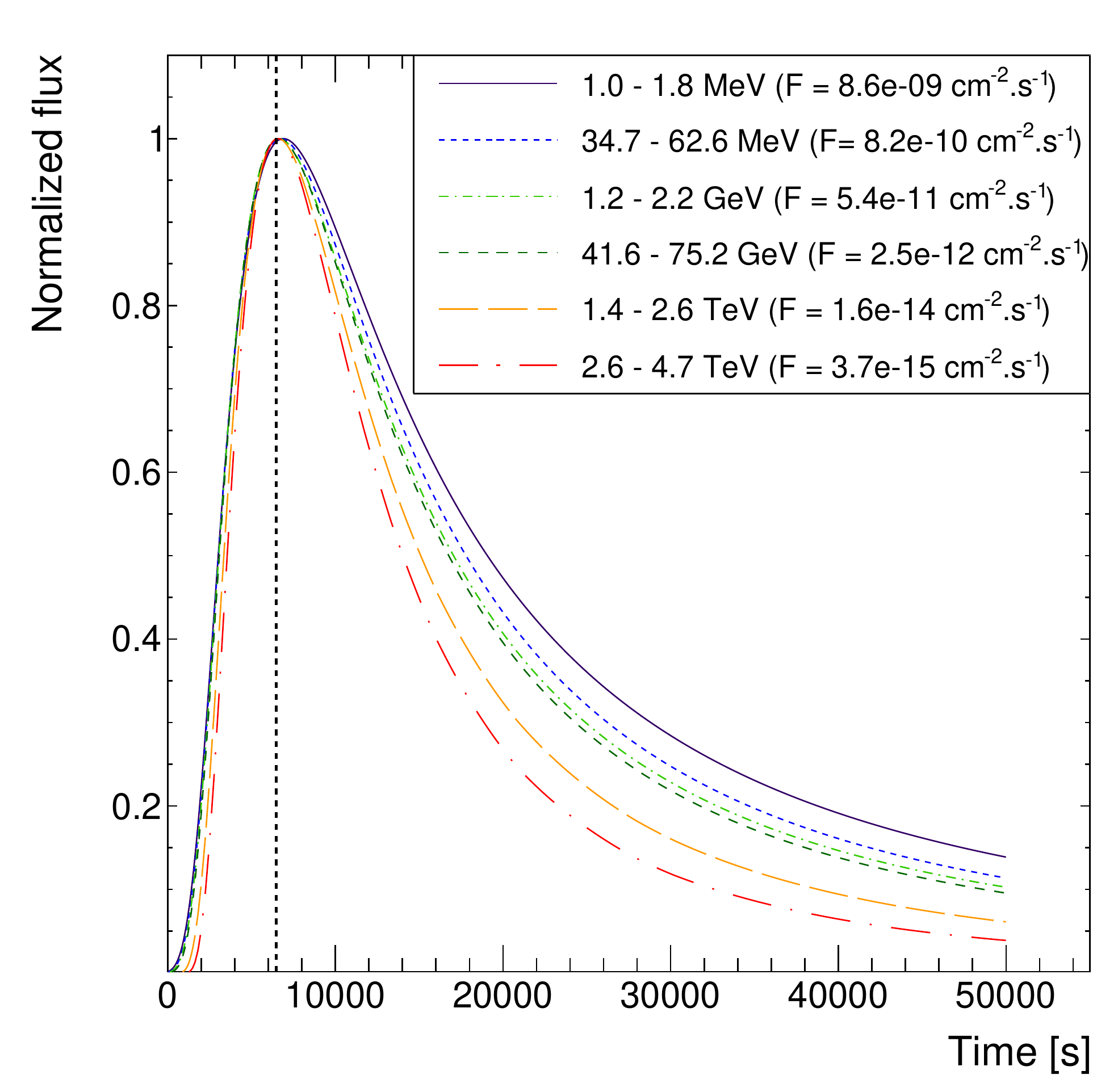}\label{fig:lc2}
        \includegraphics[width=.33\linewidth]{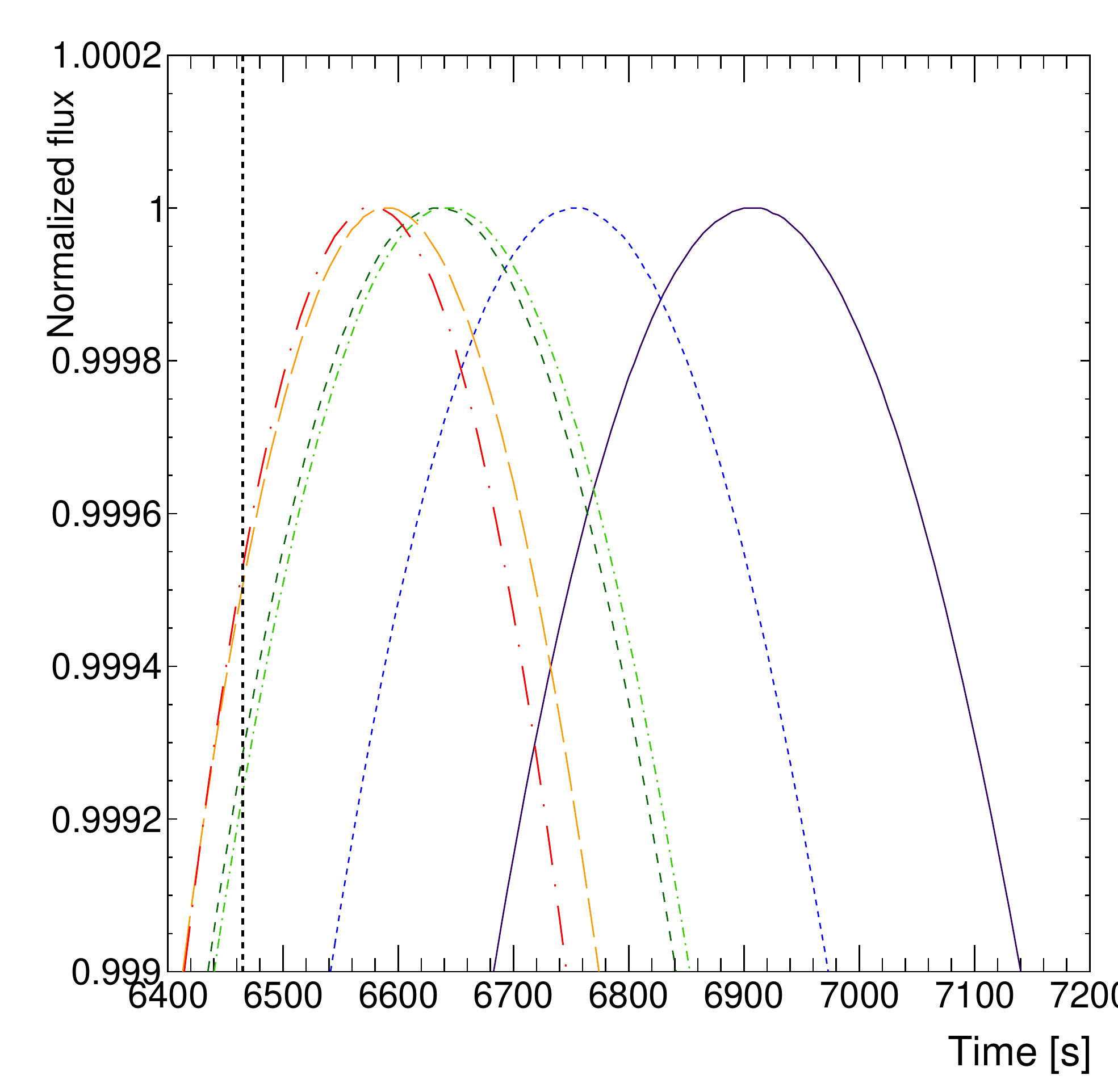}\label{fig:lc2z}
        \includegraphics[width=.33\linewidth]{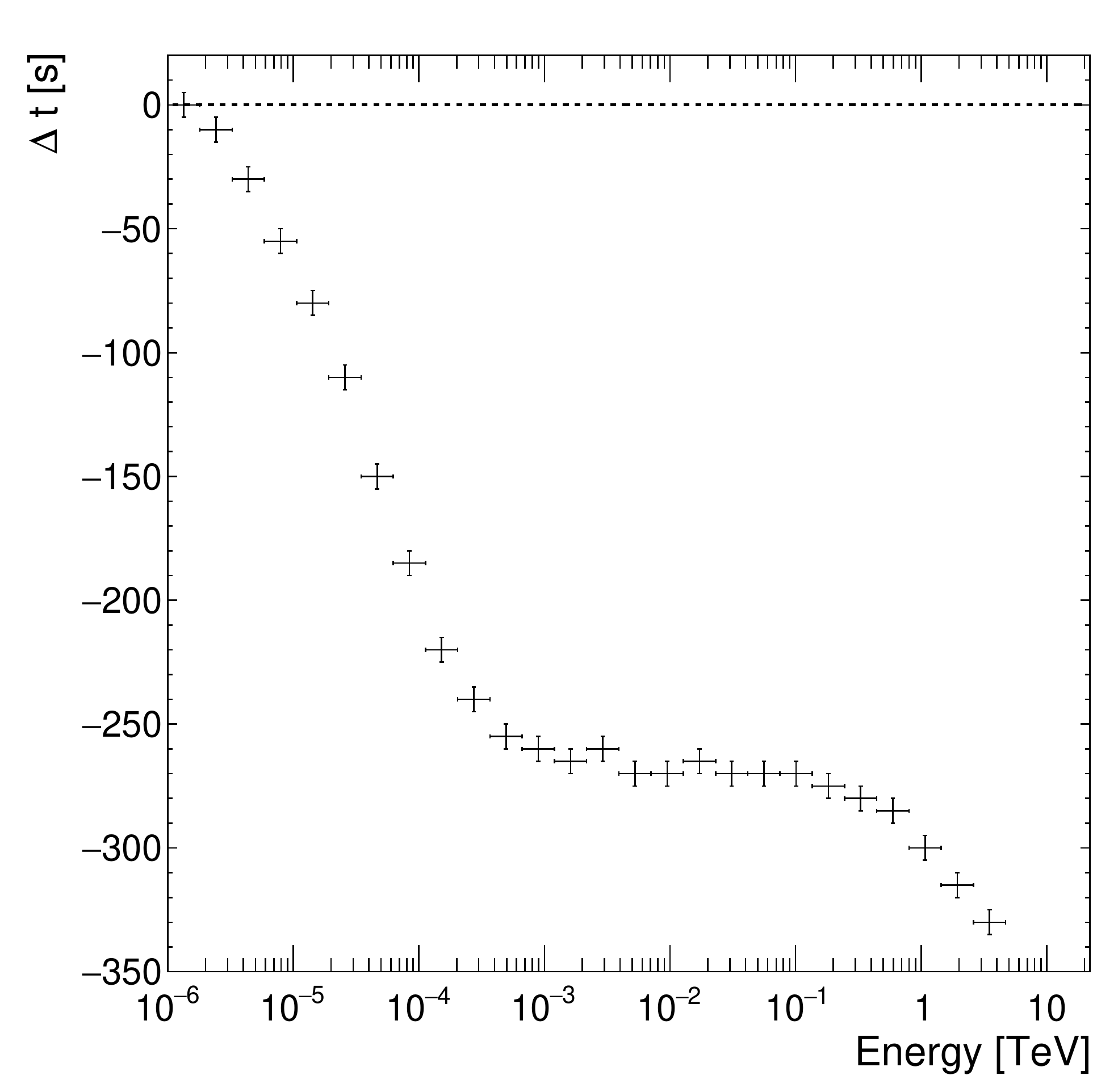}\label{fig:lag2}
\caption{Normalized light curves (left), zoom close to the light curve maxima (center), and time-delays between each light curve and the one in the range 1-1.8 MeV (right) obtained for the second reference set of parameters (i.e., with B = 90 mG). The vertical dashed line on the light curves (left and middle) corresponds to the time when electrons reach their maximum energy $\gamma_{max}$ and shows the moment when energy losses dominate over the acceleration at the energy $\gamma_{max}$.}
\label{fig:cas2}
\end{figure*}
        To compute the light curves at different energies, the SEDs are integrated over the required energy bands, each SED giving one flux point for each light curve. Examples of flare light curves are shown in Figure~\ref{fig:cas1} (left and center) for parameters of Table~\ref{tab:par_std}. In order to consider only realistic light curves that may be observed, a selection cut is applied on them. Only light curves with a maximum flux above $2 \times 10^{-15}\ \mathrm{cm^{-2}\ s^{-1}}$ are kept. The time $t_{max}$ shown on the light curve is defined from Figure~\ref{fig:sed} as the time when electrons reach their maximum energy $\gamma_{max} = max\left(\gamma_c(t)\right)$. This time indicates when the radiative cooling timescale becomes shorter than the acceleration timescale, for the most energetic electrons with an energy close to $\gamma_{max}$. For this set of parameters, the time $t_{max}$ happens after all the light curves peak, meaning that the electron acceleration timescale is still shorter than the electron cooling timescale for all energies. As a result, the decaying light curves  can only be explained by the decrease of the magnetic field. An opposite case can be defined when $t_{max}$ happens before all the light curves peak as shown in Figure~\ref{fig:cas2} (left and center). This case is obtained from the first reference set of parameters increasing the radiative cooling power. This is achieved by changing the $B_0$ value from 65~mG to 90~mG, and provides us with our second reference set of parameters throughout this paper. Conversely to the previous case, the radiative cooling  contributes to the light-curve decay, resulting in the fact that the most energetic light curves decay first.

\section{Study of time delays}
\label{sec:dt}

\subsection{Time delay determination}
\label{sec:tdd}
\begin{figure}[ht]
        \includegraphics[width=1\linewidth]{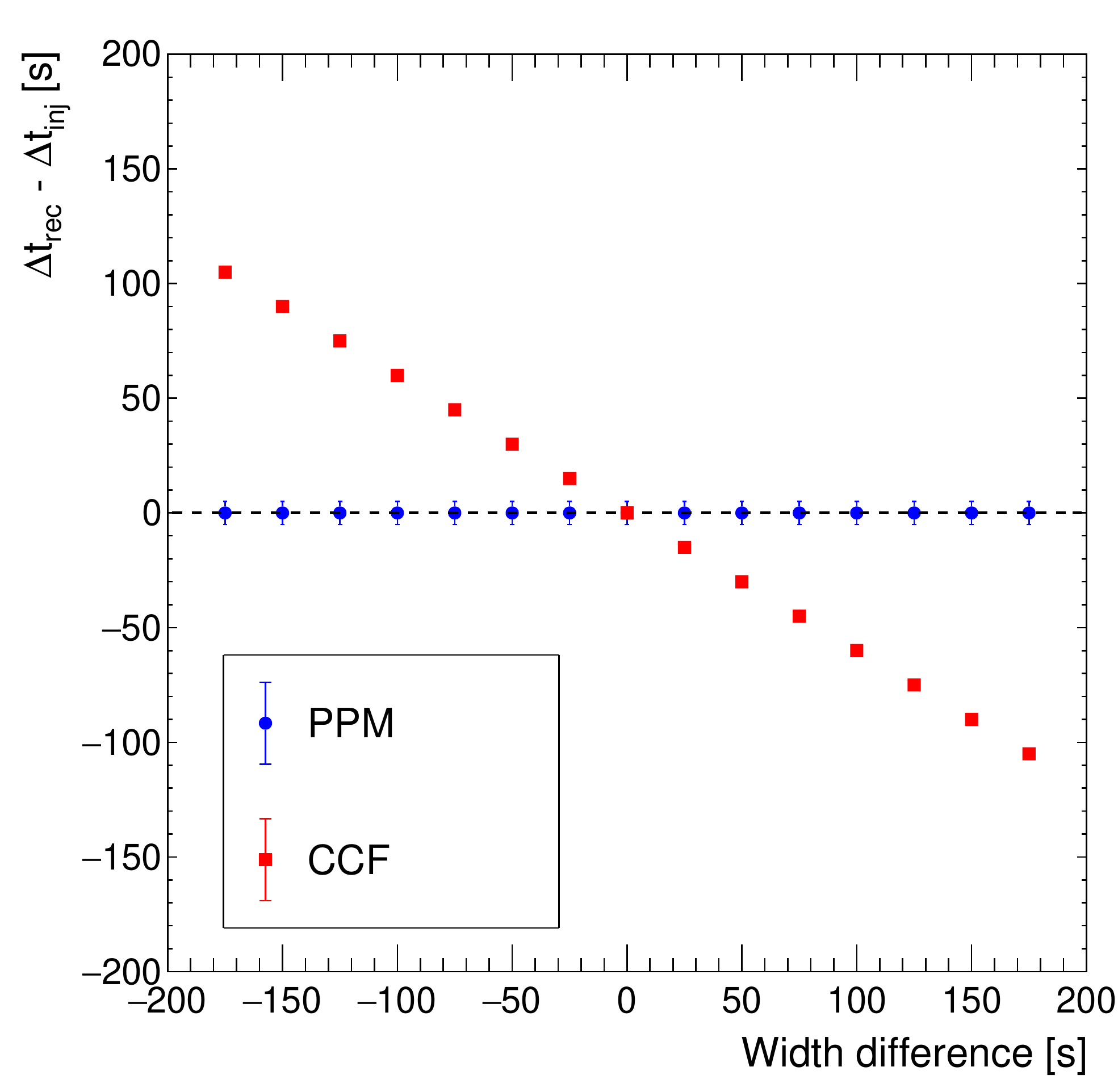}
        \caption{Time delay measured with PPM and CCF between two simulated asymmetric Gaussian light curves varying the width of one of them. The injected delay value is fixed at $\Delta t_{inj} = 800~s$. The measurement from the CCF shows a misreconstruction of the injected time delay as the width difference increases. The asymmetric Gaussian function used has $\sigma_{left} = 2500~s$ and $\sigma_{right} = 20000~s$.}
        \label{fig:method}
\end{figure}

        The time delay is determined by computing the time difference between the maximum of the light curve at the energy considered and the maximum of the lowest-energy light curve (\mbox{1.0-1.8~MeV}). A positive lag corresponds to the case when high-energy light curves peak after the lowest-energy ones. This method, later called the peak position method (PPM), is rather simplistic but provides a simple and robust estimation of the time delay, with an accuracy directly related to the time step chosen for the light curves. A cross-correlation function (CCF) from \citet{ede:1988} was also considered but was found to incorrectly reconstruct the time delay in the case of light curves with varying widths. A comparison of the PPM and CCF is shown in Figure~\ref{fig:method}. Two light curves are simulated with a similar shape as the light curves obtained from the model, approximated by an asymmetric Gaussian function. A delay of 800~s is applied to one of the light curves. The width of the lagged light curve is then varied with respect to the other one and the PPM and CCF methods are applied to reconstruct the lag. The CCF is not able to reconstruct the injected delay as the width difference increases, and more generally in case the shape of the two light curves is different. Concerning the light curves from the model, this point is further discussed in Section~\ref{sec:temp_dt}.

\subsection{Main results: two time-delay regimes}
        \label{sec:dt_regime}
        The time delays obtained for the two reference sets of parameters shown in Figure~\ref{fig:cas1} (right) and Figure~\ref{fig:cas2} (right) highlight two different time-delay regimes found with the adopted SSC flare model. These two regimes can be better characterized by considering whether the time $t_{max}$ is reached before or after the peaks of the light curves. The vertical error bars on the time-delay measurement is of 5~s and the horizontal ones for the energy correspond to the light-curve energy ranges.

        In the first case (Figure~\ref{fig:cas1}), all the light curves peak before $t_{max}$. Thus, the acceleration timescale is shorter than the cooling one when the flare decays. Electrons are still accelerated and their maximum energy still increases after the low-energy light curves peak. The decay of the flare is then induced by the decrease of the magnetic field $B(t)$. Hence, the increase of the time delay above 1~GeV is explained by the time required for electrons to be accelerated up to $\gamma_{max}$ and to emit the highest-energy photons. Conversely, below 1~GeV, the energy necessary to emit lower-energy photons is quickly reached by electrons. This leads to a decrease of the time delay, the low-energy light curves decaying due to the combined action of the energy-dependent radiative cooling and of the magnetic field decrease. In the following, this type of time delay evolution, with a decreasing time delay at low energies and an increasing one at high energies, is described as an "acceleration -driven regime"  because of the origin of the high-energy delays.

        In the second case (Figure~\ref{fig:cas2}), $t_{max}$ is reached before the peaks of the light curves, and the energy losses by radiative cooling are soon larger than the energy gains by acceleration. The decay of the flare is then mostly due to the radiative cooling. This is achieved by assuming a higher magnetic field strength initial value ($B_0 = 90$~mG) relative to the previous case, which enhances the electron radiative energy losses. As a consequence, the highest-energy light curves decay first due to a shorter cooling timescale, explaining the decreasing time delay above GeV energies. At lower energies, the cooling timescale of the electrons and the decreasing time delay appear similar to the previous case and due to the combined action of the magnetic field decrease and of the energy-dependent radiative cooling. In the following, such a case with time delays continuously decreasing are described as the "cooling driven regime" because of the main origin of the time delays above GeV energies. 

        The analysis presented in the following section confirms the existence of these two time-delay regimes over a large domain of parameters.
                
\section{Influence of model parameters on time delays}
\label{sec:param}
        \label{ref:influence}
        To investigate the impact of the model parameters on the time delays, each of them is individually varied around the reference values defined in Table~\ref{tab:par_std}, namely $B_0$, $m_\mathrm{b}$, $A_0$, $m_a$, $\delta$, $\gamma_{c, 0}$ and $n$.  All these values are chosen within a reasonable range for blazar modeling and also in the domain of validity of our model. Magnetic fields being a key driver for nonthermal blazar emission and SSC electron radiation, several values of $B_0$ are first explored and then each other parameter is investigated for $B_0 = 65$~mG and $B_0 = 90$~mG respectively. 

\subsection{Initial magnetic field strength variations}
        \label{sec:B0}
        The magnetic field directly influences the cooling timescale of electrons. Therefore, it contributes to the maximum energy  $\gamma_{max}$ reached by electrons, meaning that $B_0$ is one of the main parameters acting on the intrinsic time delays.

\begin{figure}[hb]
                \includegraphics[width=1\linewidth]{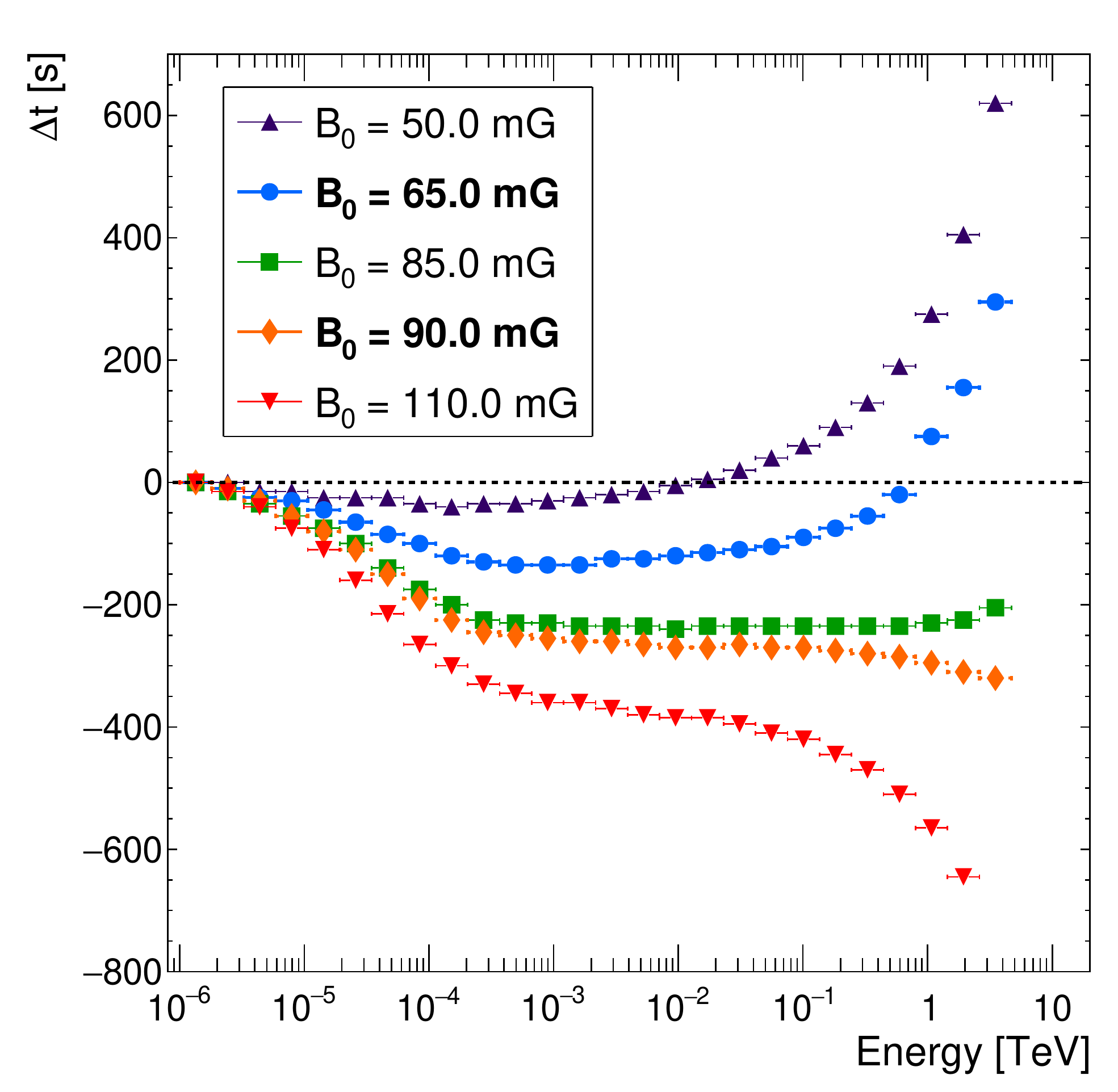}
                \caption{Time delay vs. energy for different $B_0$ values. The two cases in bold in the legend correspond to the ones discussed in Section~\ref{sec:dt_regime}. All other parameters have the values given in Table~\ref{tab:par_std}.}
                \label{fig:B0}
\end{figure}

        Several values of $B_0$ are used to evaluate the influence of the magnetic field on the time-delay evolution, ranging from $50$~mG to $110$~mG. The resulting time delay  is shown in Figure~\ref{fig:B0} as a function of energy.  In the low-energy domain, for \mbox{$E \lesssim$ 1~GeV}, the time delay does not qualitatively change with $B_0$ although it slightly decreases as $B_0$ increases due to a stronger radiative cooling. Conversely at high energies, for \mbox{$E \gtrsim$ 1~GeV}, the variation of the initial magnetic field value induces large variations of the time delay and a significant change of its behavior, with a transition zone around 80~mG from an increasing to a decreasing phase when $B_0$ increases. Within this transition, some cases show an almost constant time delay relative to the MeV light curve, which means that they present just very small or even no relative time delay in the limited energy range from 1 GeV to 1 TeV. 

\subsection{Magnetic field temporal index}
\begin{figure}[ht]
                \includegraphics[width=.49\linewidth]{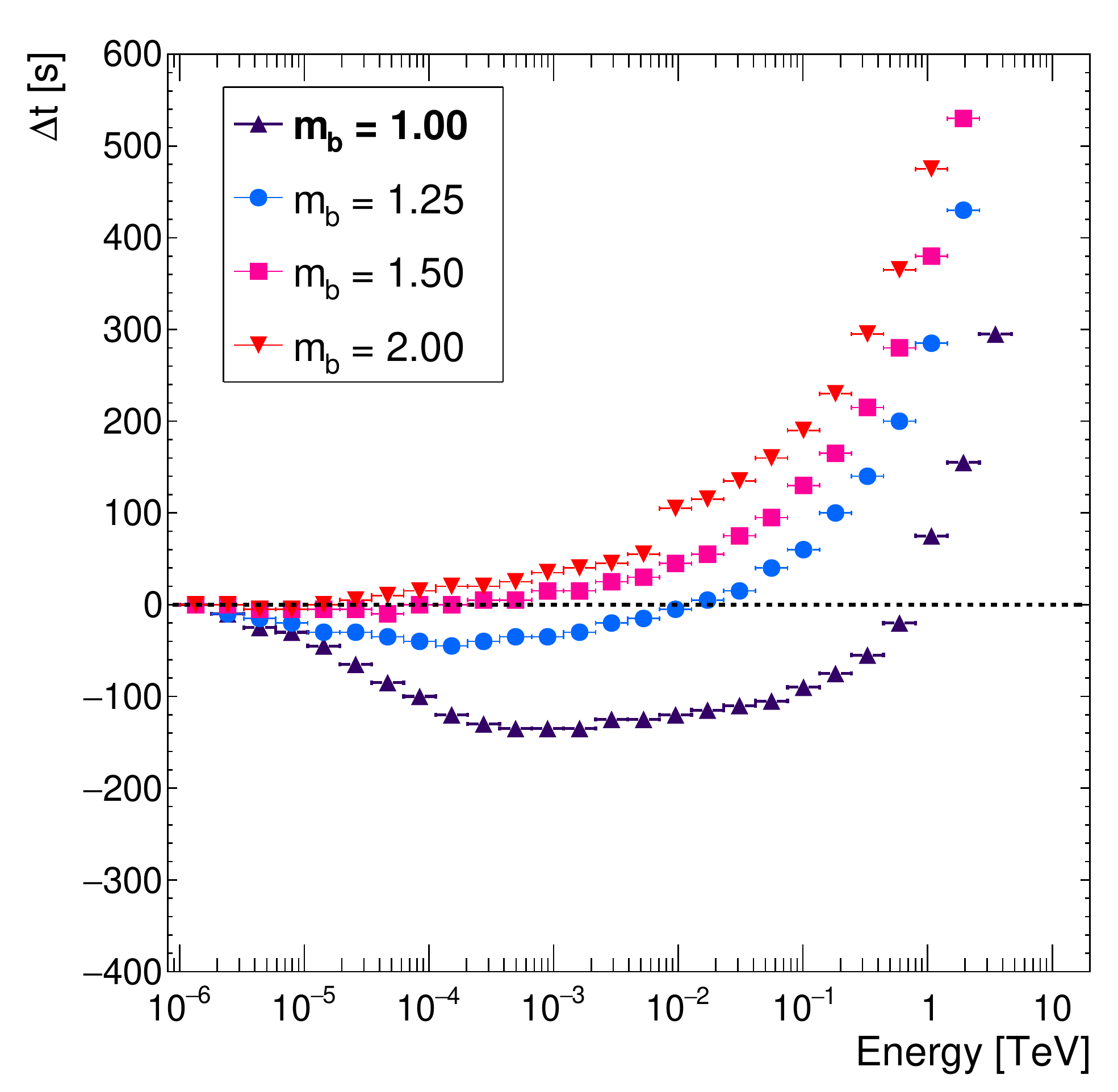} \hfill
                \includegraphics[width=.49\linewidth]{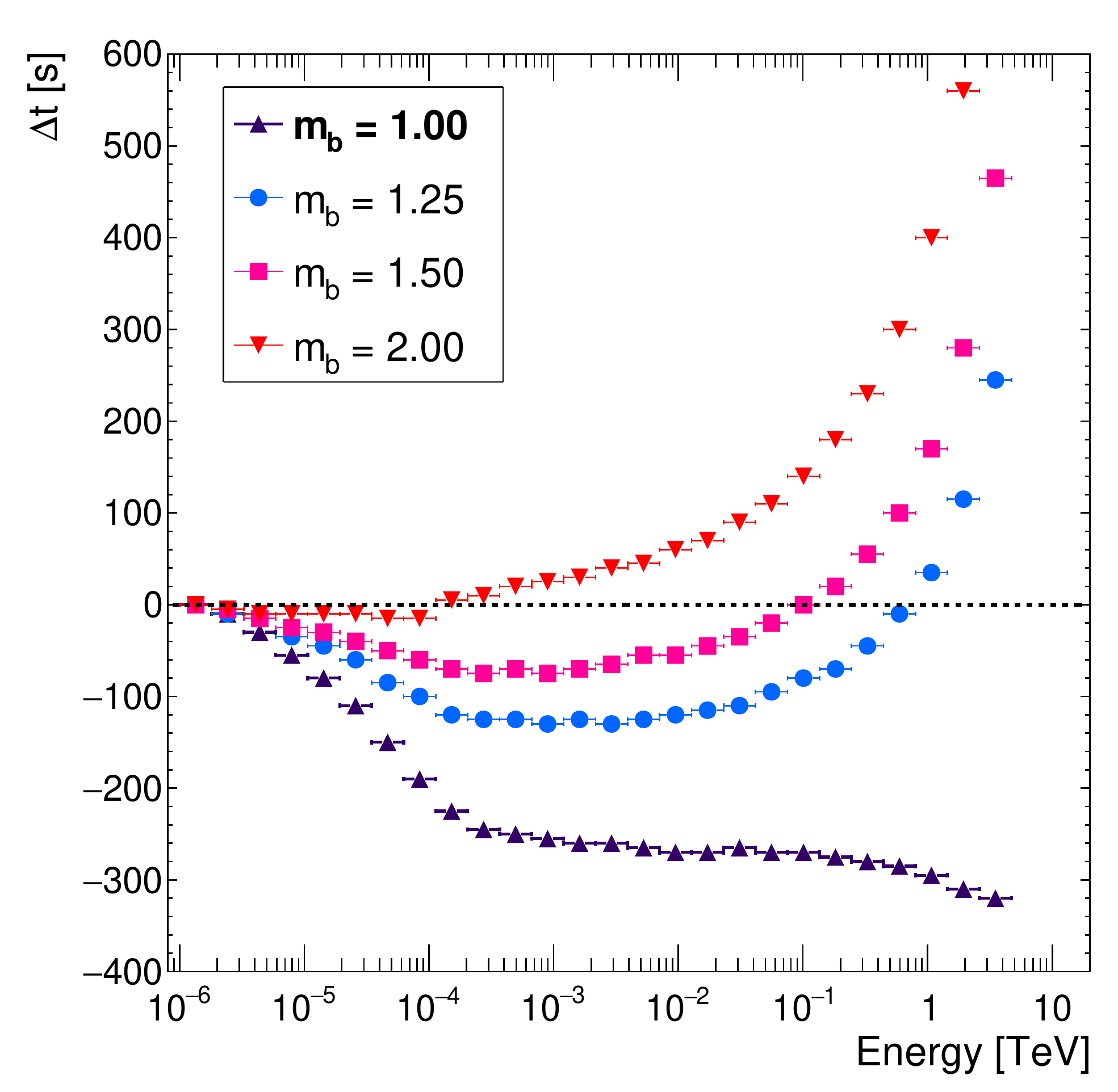}
                \caption{Time delay vs. energy for different magnetic field evolution index $m_\mathrm{b}$ with $B_0 = 65$~mG (left) and  $B_0 = 90$~mG (right). All other parameters are unchanged (Table~\ref{tab:par_std}). The two cases in bold in the legend correspond to the situation discussed in Section~\ref{sec:dt_regime}.}        
                \label{fig:mb}
        \end{figure}

        As defined in Equation~\ref{eq:Bt}, the magnetic field strength is assumed to decrease over time with an index $m_\mathrm{b}$. A value $m_\mathrm{b} = 1$ was found adequate to reproduce flares observed in Mrk 421 within the SSC scenario adopted here \citep{kat:2003}, and  $m_\mathrm{b} = 2$ corresponds for instance to the case of magnetic flux conservation in a blob propagating along a quasi-conical jet. Here, the index $m_\mathrm{b}$ is varied from 1 to 2 and time delays are shown for the two reference cases in Figure~\ref{fig:mb}.
                        
         Starting from an acceleration-driven regime in the case where $B_0 = 65$~mG, an overall increase of the time delay value is observed. For \mbox{$E \gtrsim$ 1~GeV} the time delay evolution does not qualitatively change and lags always increase. For \mbox{$E \lesssim$ 1~GeV}, the time delay decreases less and less as $m_\mathrm{b}$ increases and finally evolves towards only increasing time delay. This transition is explained by the rapid decrease of the magnetic field which induces low-energy flares quickly decaying while electrons are still accelerating. As a consequence, the highest -energy light curves peak at later times  due to the time needed for the electrons to reach high $\gamma$ values.   
                        
        Starting from a cooling driven regime in the case where $B_0 = 90$~mG, increasing $m_\mathrm{b}$ induces a transition to the acceleration-driven regime. In those cases, the rapid decrease of $B(t)$ reduces the radiative cooling power. Hence the acceleration time scale becomes shorter than the radiative cooling timescale, thus shifting $t_{max}$ to later times. When $m_\mathrm{b}$ is larger, the highest-energy light curve peaks later while the faster magnetic field decrease induces the lowest-energy light curves to decay earlier.
                        
\subsection{Doppler factor variations}
\begin{figure}[ht]
                \includegraphics[width=.495\linewidth]{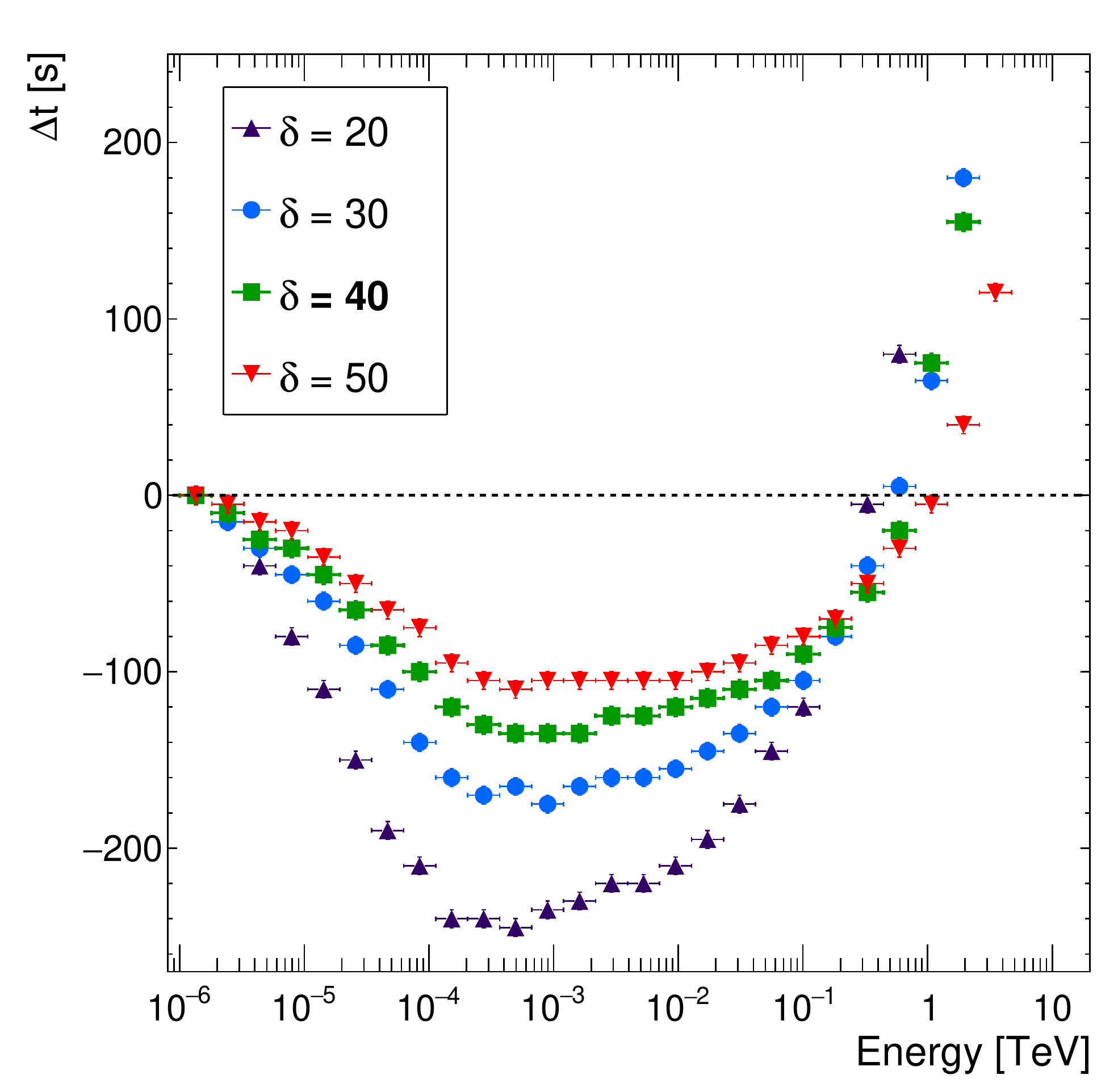} \hfill
                \includegraphics[width=.495\linewidth]{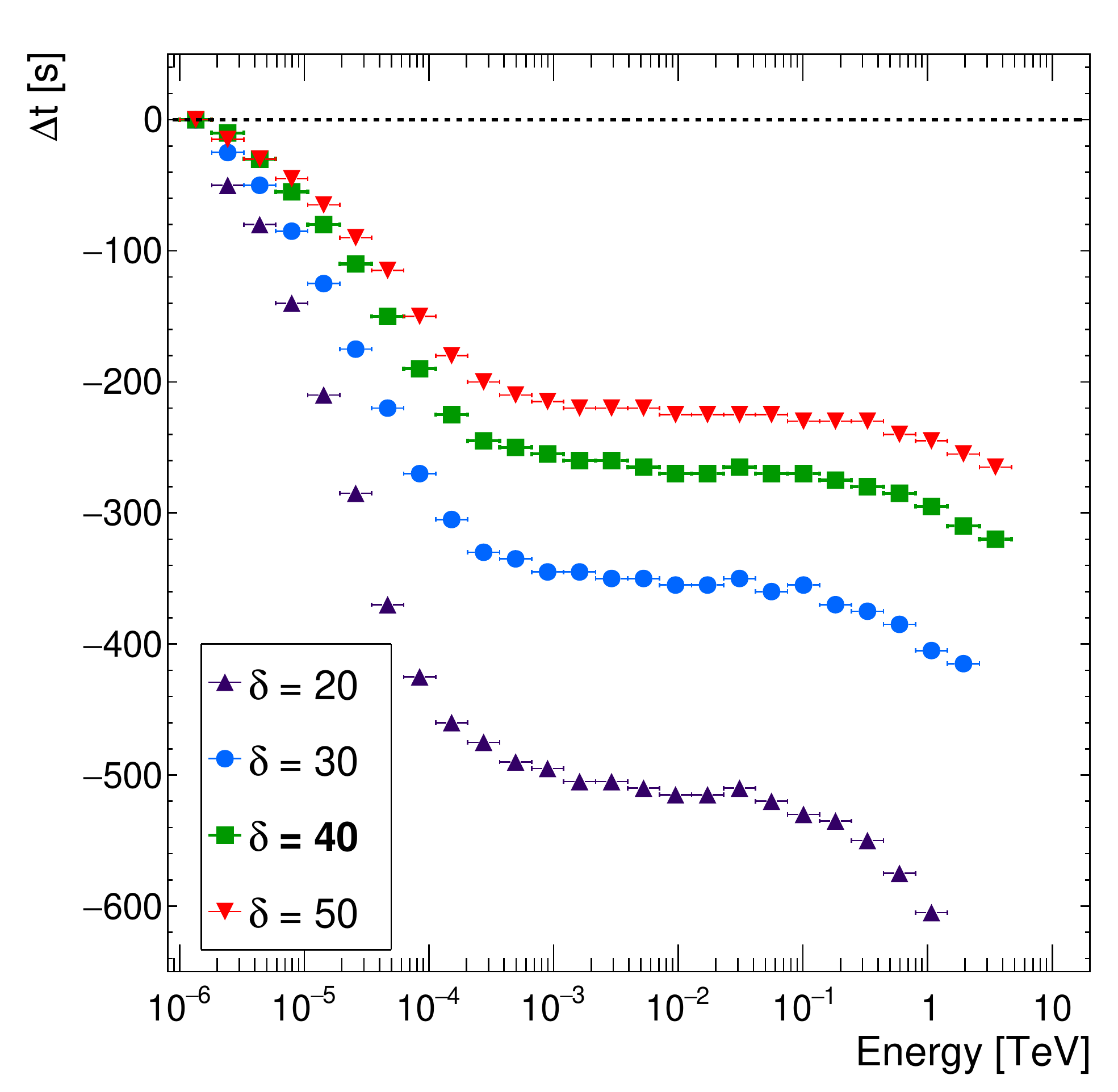}
                \caption{Time delay vs. energy for different Doppler factor $\delta$ values with $B_0 = 65$~mG (left) and  $B_0 = 90$~mG (right).  All other parameters are unchanged (Table~\ref{tab:par_std}). The two cases in bold in the legend correspond to the situation discussed in Section~\ref{sec:dt_regime}.}
                \label{fig:dop}
        \end{figure}

                        A modification of the Doppler factor $\delta$ does not change the temporal evolution of the source in its own frame and simply implies a change on the Doppler boosting effect for the observer leading to variations of the observed variability due to time contraction as well as variations of the observed flux and energy. The energy-dependent time delay for different $\delta$ ranging from 20 to 50  is shown in Figure~\ref{fig:dop} for the two regimes. The dominant effect on the time-delay evolution is the time contraction, inducing smaller observed time delays as $\delta$ becomes larger. In reality, the time delays obtained for different $\delta$ values appear almost proportional to each other by the ratio of their Doppler factor, as shown for instance for $B_0 = 90$~mG by the delay at 1~TeV for $\delta = 40$ and  $\delta = 20$ which gives \mbox{$\Delta t_{\delta=40}(1~\mathrm{TeV}) \simeq 293$~s} and \mbox{$\Delta t_{\delta=20}(1~\mathrm{TeV}) \simeq 600$~s}, with a ratio approximately equal to the Doppler factor ratio. However, it can also be noted that the variation of the maximum energy considered for the time-delay computation is an effect of the Doppler boosting on the apparent  flux value. This is a consequence of our choice to neglect light curves with flux values below $2 \times 10^{-15}\ \mathrm{cm^{-2}\ s^{-1}}$. Also, the energy shift due to Doppler boosting can lead to a change of the time delay sign which is not expected from the time contraction alone as seen for $B_0 = 65$~mG at $E \approx 1$~TeV.

\subsection{Acceleration parameter variations}
        The acceleration term, defined in Equation~\ref{eq:Cacc}, depends on two parameters, namely the initial amplitude $A_0$ and the evolution index $m_a$. Varying the acceleration parameters modifies the electron acceleration time scale and thus the time $t_{max}$ with respect to the flare maxima. This can in turn induce a transition between the two time-delay regimes.
                        
\begin{figure}[ht]
                \includegraphics[width=.495\linewidth]{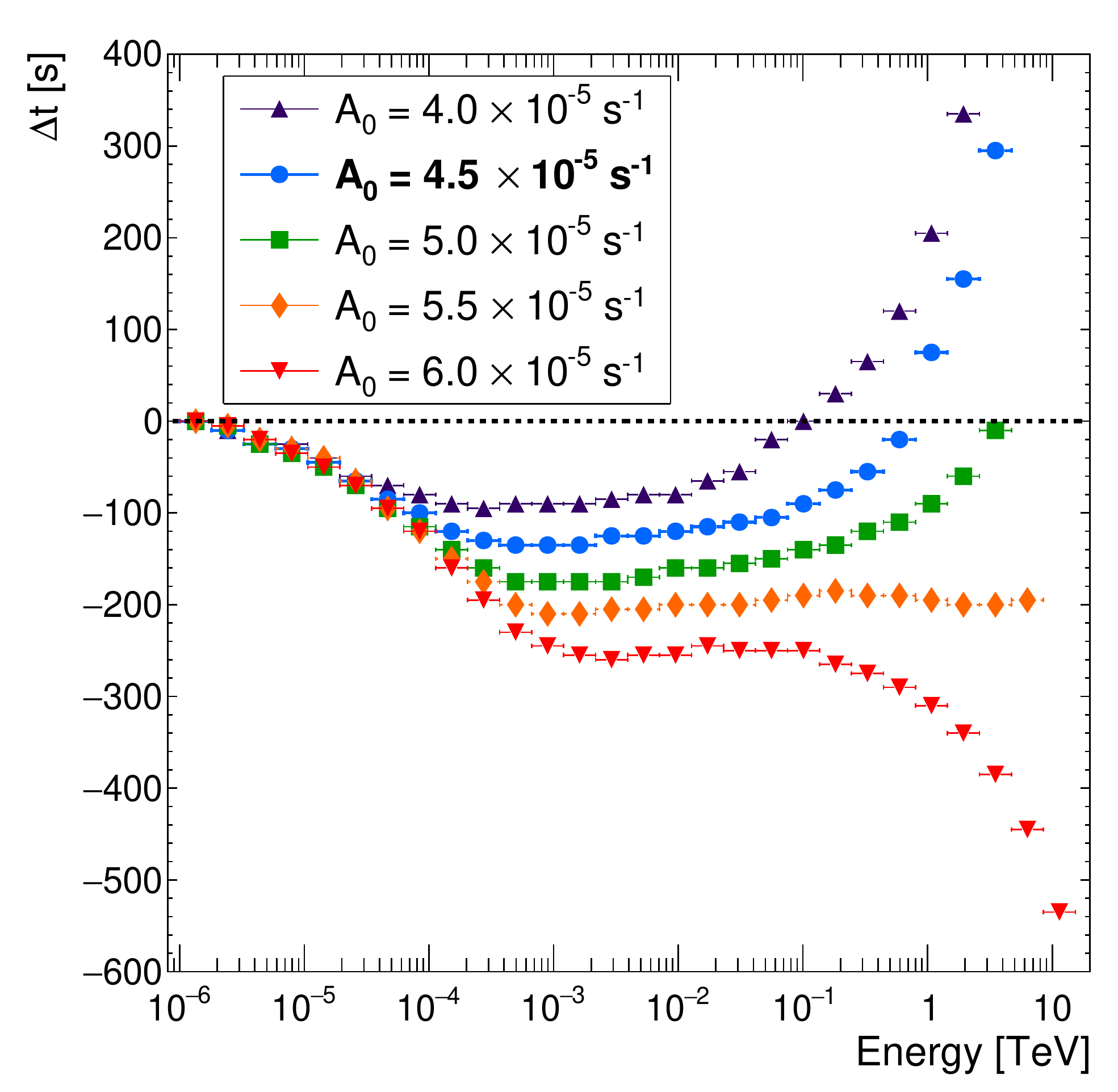}
                \includegraphics[width=.495\linewidth]{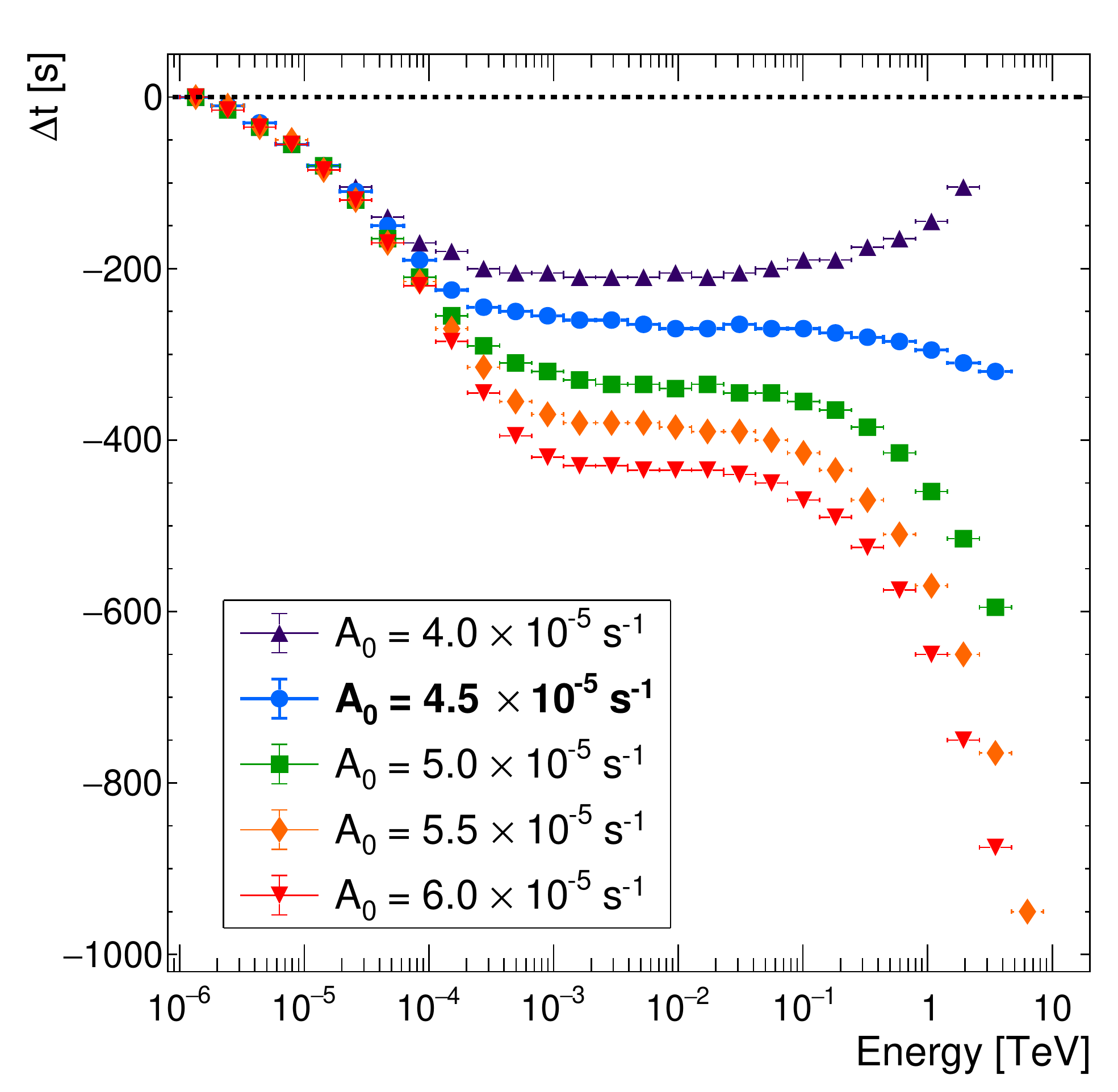}
                \caption{Time delay vs. energy for different acceleration amplitude $A_0$ values with $B_0 = 65$~mG (left) and  $B_0 = 90$~mG (right). All other parameters are unchanged (Table~\ref{tab:par_std}). The two cases in bold in the legend correspond to the situation discussed in Section~\ref{sec:dt_regime}.}
                \label{fig:A0}
\end{figure}

        The time delay obtained for different $A_0$ values ranging from \mbox{$4.0\times10^{-5}$~s$^{-1}$} to \mbox{$6.0\times10^{-5}$~s$^{-1}$} is shown in Figure~\ref{fig:A0}. In the case where $B_0=65$~mG, decreasing $A_0$ increases the time delays for the highest energies. The acceleration power is weaker and high-energy electrons need more time to reach $\gamma_{max}$. In addition, the maximum electron energy $\gamma_{max}$ is smaller, implying a weaker cooling effect which shifts $t_{max}$ to later times. With larger $A_0$ values, a transition between the two regimes occurs, leading to a cooling-driven regime. The acceleration is then stronger involving a higher $\gamma_{max}$ value, inducing a stronger cooling effect for the most energetic electrons. Hence, increasing $A_0$ brings electrons much faster to higher $\gamma_{max}$ values which then quickly suffer from intense radiative energy losses, entering into the cooling-driven regime.
                        
        Conversely, starting from a cooling-driven regime with $B_0 = 90$~mG, the opposite situation occurs with a transition to the acceleration-driven regime when $A_0$ decreases. Small $A_0$ values induce longer acceleration timescales and lower $\gamma_{max}$ values, the cooling power at $\gamma_{max}$ becomes  weaker, electrons need a long time to reach their maximum energy, the $t_{max}$ is shifted to later times, and a transition occurs to the acceleration-driven regime  when $A_0$ is small enough. On the other hand, larger $A_0$ values imply a shorter acceleration timescale, $\gamma_{max}$ is reached at earlier times, and the cooling power becomes stronger. Then the highest-energy light curves reach their peaks much earlier than the reference case, which explains the negative time delays.

\begin{figure}[ht]
                \includegraphics[width=.495\linewidth]{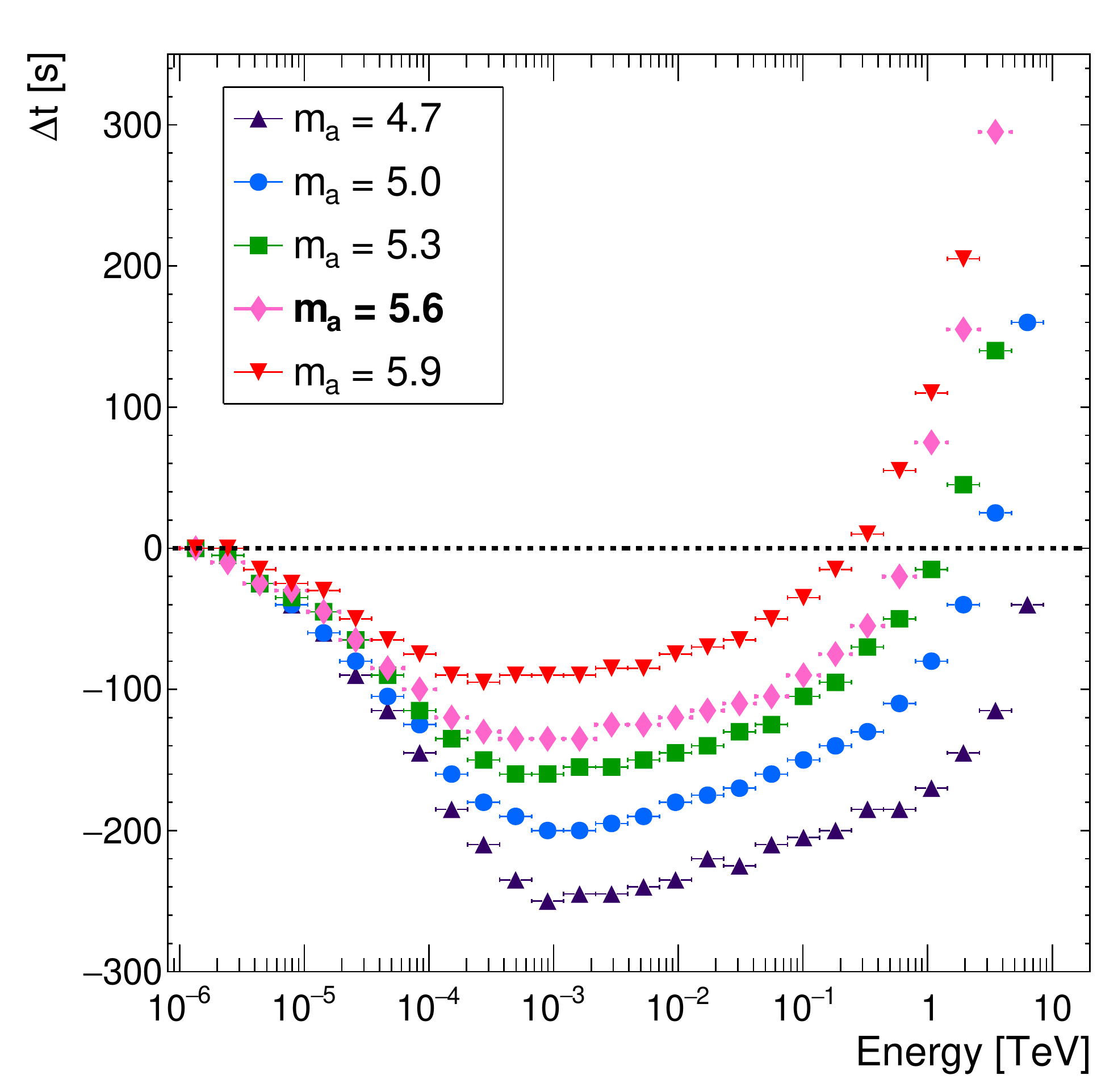} \hfill
                \includegraphics[width=.495\linewidth]{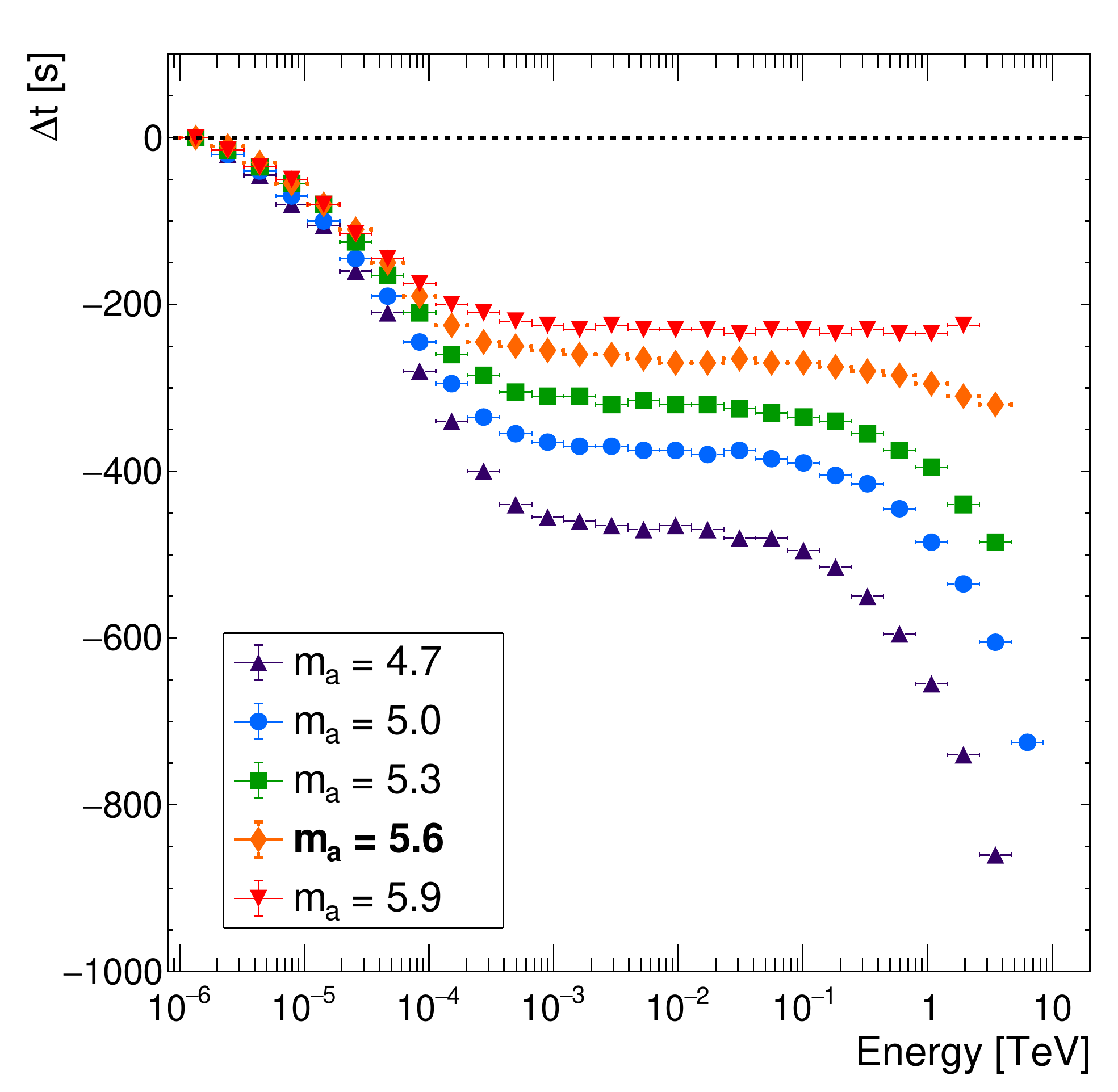}
                \caption{Time delay vs. energy for different acceleration evolution $m_a$ values with $B_0 = 65$~mG (left) and  $B_0 = 90$~mG (right). All other parameters are unchanged (Table~\ref{tab:par_std}). The two cases in bold in the legend correspond to the situation discussed in Section~\ref{sec:dt_regime}. }
                \label{fig:a}
\end{figure}

        Similar reasoning applies to the acceleration evolution index parameter $m_a$ (Figure~\ref{fig:a}), varied from 4.5 to 5.9. Increasing $m_a$ induces longer acceleration timescales. However, the variation of $m_a$ does not lead to a significant change of regime. For $B_0 = 90$~mG, a hint of transition is observed for large $m_a$ values when the acceleration is weaker. For $B_0 =  65$~mG, the overall delay decreases toward negative values for a stronger acceleration when $m_a$ decreases. 

\subsection{Initial electron distribution index variations}
        The initial electron spectrum assumed in the present flare scenario follows a power law function with a  high-energy cut-off (Equation~\ref{eq:N0}). The initial electron density $K_0$  is only a scaling parameter and does not affect the time evolution.  In the transfer equation, modifying $n$ does not change the balance between electron acceleration and cooling effects and in practice $\gamma_{max}$ and $t_{max}$ remain at the same values when $n$ changes. However, the initial electron spectrum index $n$ impacts the ratio of low- to high-energy electrons, with a higher proportion of lower-energy electrons for high $n$ values.
                        
\begin{figure}[ht]
                \includegraphics[width=.495\linewidth]{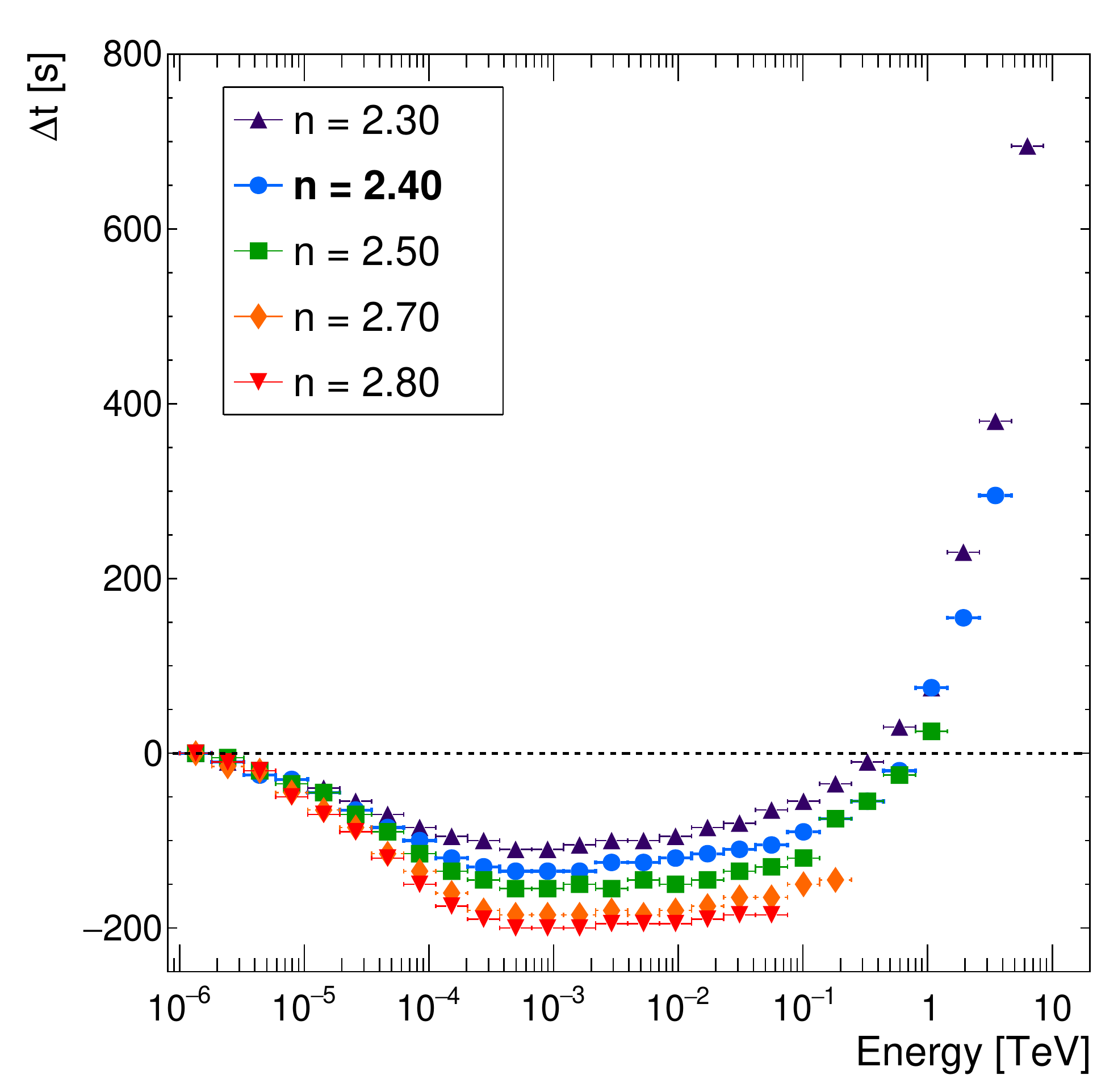} \hfill
                \includegraphics[width=.495\linewidth]{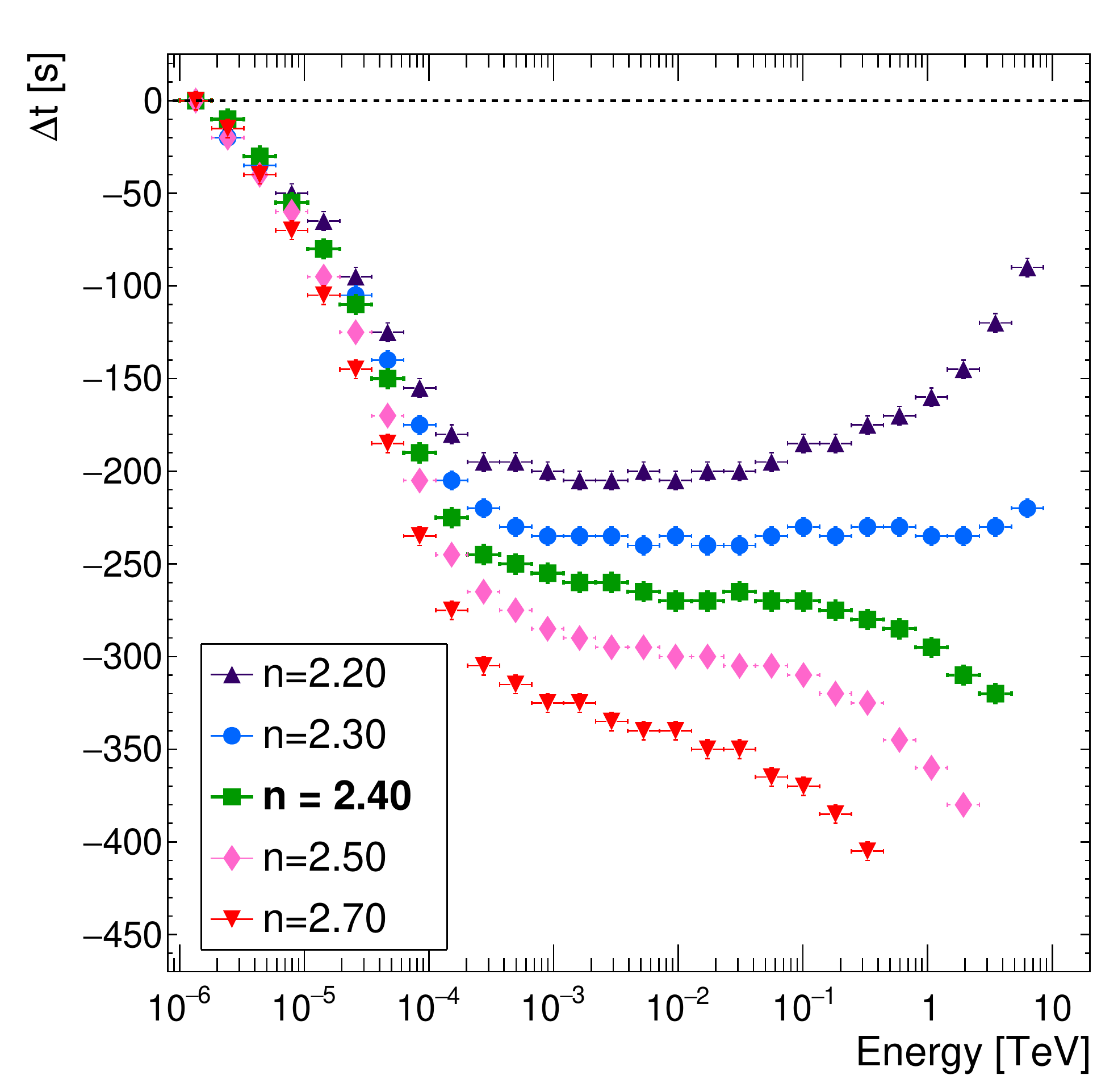}
                \caption{Time delay vs. energy for different initial electron index $n$ values with $B_0 = 65$~mG (left) and  $B_0 = 90$~mG (right). All other parameters are unchanged (Table~\ref{tab:par_std}). The two cases in bold in the legend correspond to the situation discussed in Section~\ref{sec:dt_regime}.}
                \label{fig:n}
        \end{figure}

        The time delays obtained with values from $n = 2.2$ to $n = 2.8$ are presented in \mbox{Figure~\ref{fig:n}}. For $B_0 =  90$~mG, starting from a cooling-driven regime, a transition occurs to the  acceleration-driven regime when $n$ decreases. Such a transition can be easily explained because the flare is globally shorter when $n$ is smaller while $t_{max}$ remains the same whatever the value of $n$. This is the consequence of the fact that the electron population is on average more energetic for smaller $n$,  therefore inducing  light curves which peak at earlier times. The transition occurs when $n$ is small enough to produce light curves peaking earlier than $t_{max}$.  For $B_0 = 65$~mG, the variations of the time delay are small  and do not really highlight the influence of the parameter $n$, however the evolution of the time delay follows the same behavior. A smaller $n$ value induces an overall shorter flare, with light curves peaking at earlier times  because they are produced by a more energetic electron population. The 
maxima of the light curves become shifted at earlier times than $t_{max}$ thus inducing larger delays for the highest-energy light curves. 

\subsection{Energy break of electron distribution evolution}
\begin{figure}[ht]
                \includegraphics[width=.495\linewidth]{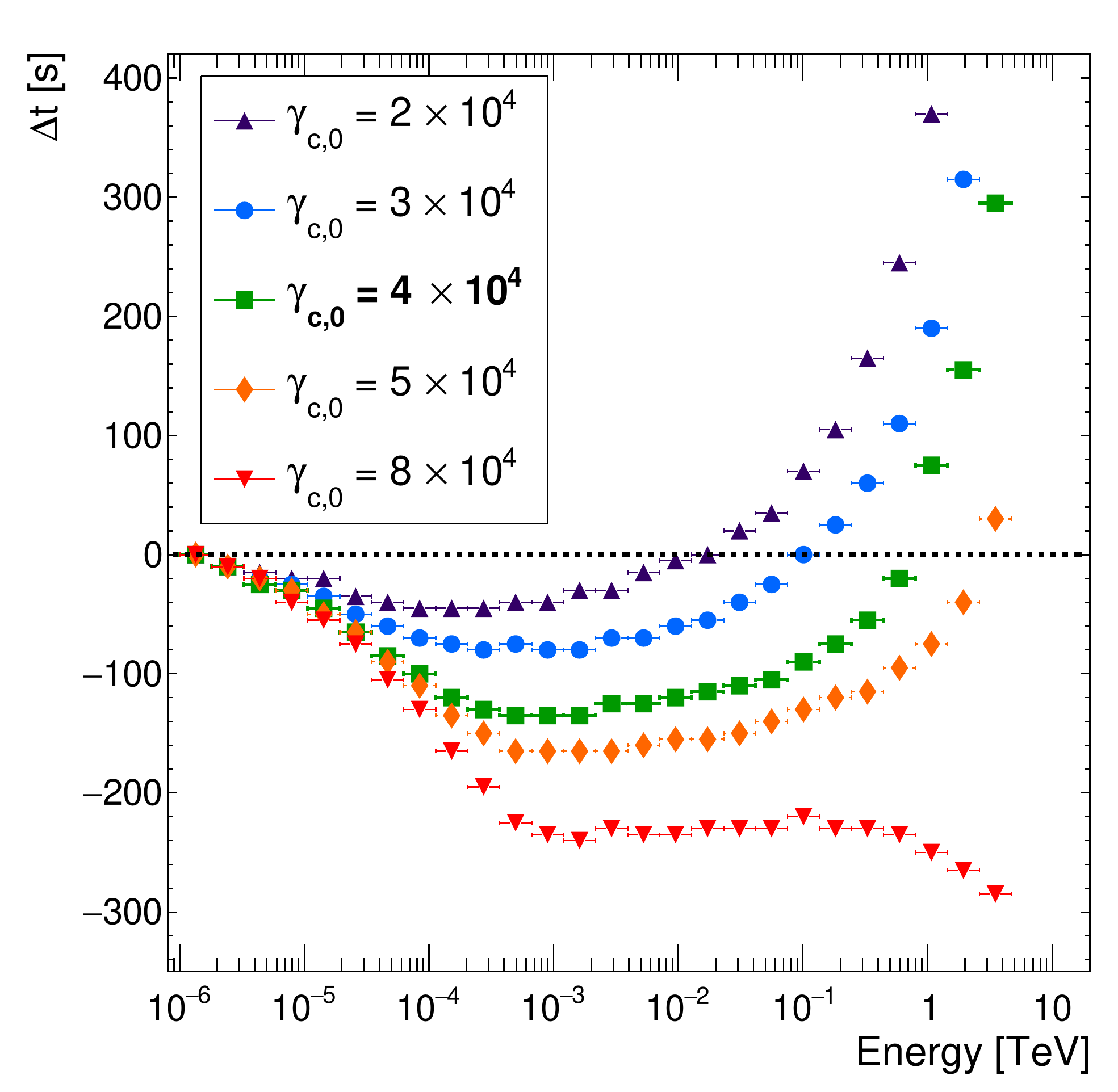} \hfill
                \includegraphics[width=.495\linewidth]{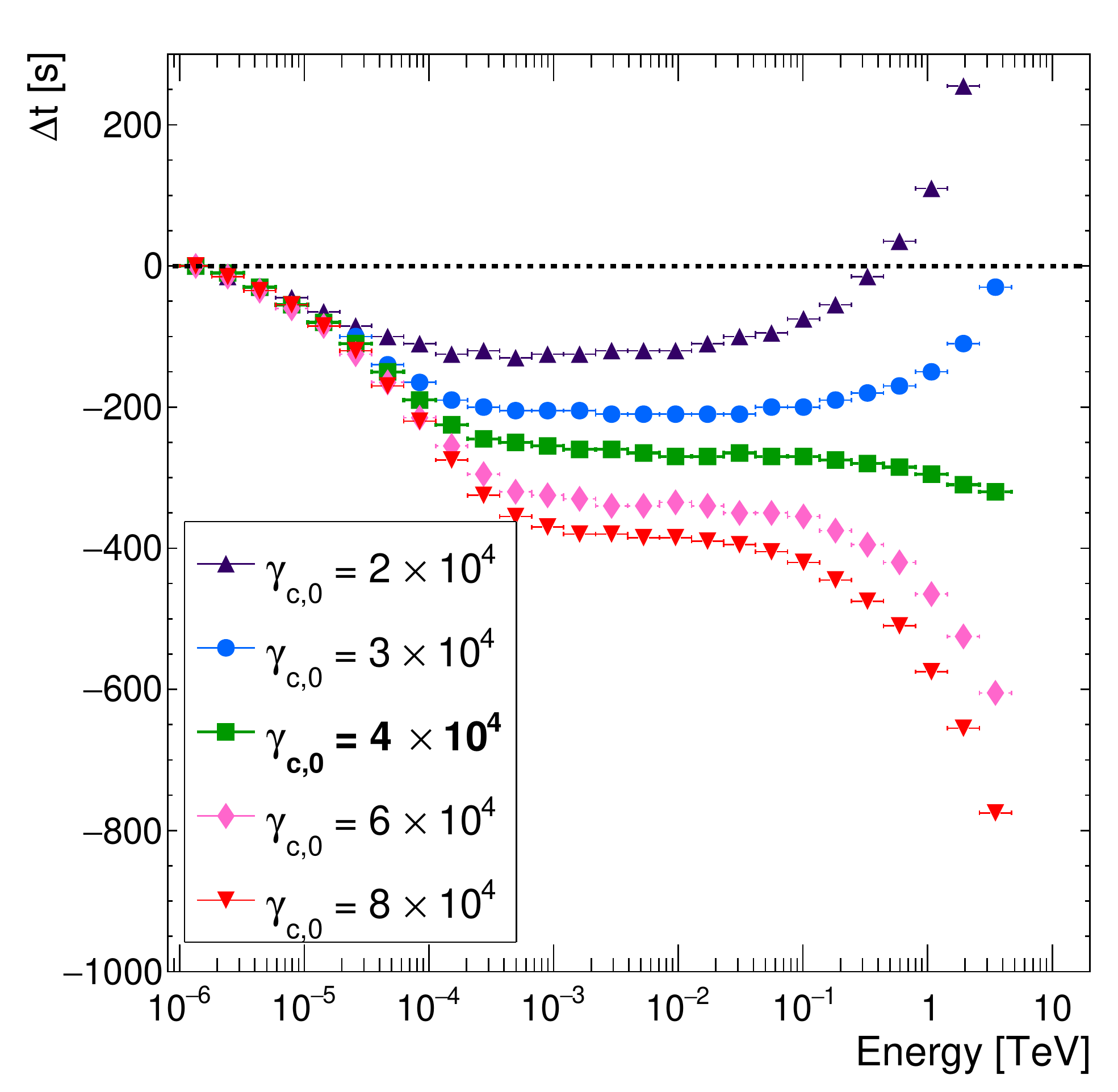}
                \caption{Time delay vs. energy for different initial maximum energy for electron $\gamma_{c,0}$ values with $B_0 = 65$~mG (left) and  $B_0 = 90$~mG (right).  All other parameters are unchanged (Table~\ref{tab:par_std}). The two cases in bold in the legend correspond to the situation  discussed in Section~\ref{sec:dt_regime}.}
        \label{fig:gmax}
\end{figure}

        The energy cut-off $\gamma_{c,0}$ defines the maximum energy of electrons at the starting time $t_0$ (Equation~\ref{eq:N0}). Using lower values of $\gamma_{c, 0}$ than $4\times10^{4}$ increases the time needed for electrons to reach their highest energy $\gamma_{max}$. For higher $\gamma_{c, 0}$ values, electrons are quickly accelerated to high $\gamma$ values and $\gamma_{max}$ become larger leading to a shorter cooling timescale. The resulting time delays for $\gamma_{c,0}$ ranging from $2 \times 10^4$ to $8 \times 10^4$ are shown in \mbox{Figure~\ref{fig:gmax}}  for the acceleration- and cooling-driven cases.
                         
        For $B_0 = 65$~mG, decreasing $\gamma_{c,0}$ leads to an increase of the time delay values for all energies. Indeed, starting from less energetic electrons implies that they need more time to be accelerated up to $\gamma_{max}$, thus shifting the highest-energy light curves to later times. For larger $\gamma_{c,0}$ values, the electron population is more energetic at $t_0$ and reaches a larger $\gamma_{max}$ within a smaller time $t_{max}$. In this situation, the cooling timescale becomes shorter at $\gamma_{max}$ leading to a transition from the acceleration- to the cooling-driven regime. For \mbox{$B_0 = 90$~mG}, a similar behavior appears. 
        
\section{Discussion and astrophysical issues}
        With the flare model presented here, all the cases investigated reveal the presence of an energy-dependent intrinsic time delay at gamma-ray energies. Two distinct regimes, referred to as ``cooling driven'' and ``acceleration driven''  are found for the time delays, corresponding to the mechanism driving the electron evolution when the light curves peak and the flare starts to decay. In addition, some specific cases corresponding to the transition between the two regimes show no delay between roughly 100 GeV and a few TeV. Adiabatic effects due to the expansion of the emitting zone can affect the quantitative values of time delays but do not qualitatively modify the global picture. When added in the differential equation describing the evolution of electrons (Equation~\ref{eq:equadiff}), they contribute to reducing the acceleration term, therefore pushing the system towards the acceleration-driven regime, which will be reached or not depending on the given set of parameters. 
        
\subsection{Observational constraints on the model}
         The information on the energy-dependent time delay can be used in order to constrain the model parameters. Indeed, if the evolution of the measured time delay corresponds to one of the two regimes, the other one is obviously ruled out. Clearly, the time delay is a new observable which can be used to constrain the modeling of blazars.
                
                 For instance, the observation of a flare from the blazar Markarian~501 in 2005 \citep{Albert2007} revealed a nonzero time delay in the VHE range, the unique case of detection of a time delay at VHE from a blazar. The authors found a time delay  increasing with the energy which appears to correspond to an  acceleration-driven regime. This suggests a qualitative scenario with a flare initiated through a sudden shock acceleration or magnetic reconnection in the emitting zone, immediately followed by a mechanism inducing the flare decay such as the decrease of the magnetic field (or adiabatic expansion). In parallel, acceleration processes are still more efficient than the radiative cooling at the highest energies and thus ensure the observed acceleration-driven regime.
                 
                However, most of the blazar flare observations do not show any significant time delays. One example is provided by the exceptional flare of PKS 2155-304 observed by H.E.S.S.  in 2006. A CCF was applied to the light curves \citep{aha:2008} between 400 and 800~GeV and above 800~GeV, and no significant delay was found in the data. The reason for such missing time-lags is not clear, because basic SSC flare models such as the one presented here predict significant intrinsic delays. Obviously this can be due to the limited number of observed blazar TeV flares, the poor time coverage and time accuracy, and the limited spectral range of the present data sets. The next generation of gamma-ray instruments should clarify this issue by measuring significant time delays and providing new and precise quantitative constraints on VHE flare models. Indeed, a flare which can only decay through radiative cooling leads inevitably to a cooling-driven regime. Therefore, for example,  if future observations of blazar flares do not reveal the presence of time delays decreasing with the energy at VHE, the simplest scenarios with a fast acceleration (or injection) followed by radiative cooling in otherwise constant emitting zone and magnetic field could be excluded. Alternatively, it is also possible that basic SSC scenarios do not describe blazar flares in full details. If time delays are not at all observed at the VHE, then the majority of flares may occur in a specific domain of parameters corresponding to the transition zone between the two time-delay regimes we have identified. This would suggest a physical link and a fine-tuning between acceleration and cooling processes in the global evolution of the flares. 

\subsection{Focus on time delays at very high energies}
        For the search of LIV signatures with IACT, studies are performed in the VHE range only. The intrinsic delays obtained in Sections~\ref{sec:dt} and \ref{sec:param} are therefore re-calculated here relative to a reference light curve at higher energies, in the range from $42$ to $74$~GeV which is now used as the zero origin for the delays. Comparing the energy dependency of the intrinsic time delays with the LIV ones can then provide direct constraints on specific QG models since for instance some models produce only one specific type of time delay such as \citet{ame:1997} or \citet{ell:2000} with positive,  linear, energy-dependent delays.
        
        To quantify the energy dependency of the intrinsic delays, they are adjusted with a power law function similar to the one used for LIV studies, namely
        
        \begin{equation}
                \Delta t = \xi \times \left(E_i^{\alpha} - E_0^{\alpha}\right),  \label{eq:fit_nrj}
        \end{equation}          
        where $\alpha$ is the energy dependency index, $\xi$ the amplitude of the delay in $\mathrm{s\ TeV^{-\alpha}}$, and $E_0$ the midpoint of the energy range of the reference light curve (58.5~GeV). The $\alpha$ and $\xi$ parameters obtained from fitting the time delay for the cases with different $B_0$ values are shown in \mbox{Table~\ref{tab:alpha_B}} and for all other cases in \mbox{Table~\ref{tab:alpha_all}} of the Appendix. At the transition between the two regimes, $\alpha$ cannot be evaluated since there is no significant time delay. In addition, some cases are not able to produce a significant flare above 250~GeV, implying a poor energy coverage and preventing any estimation of the $\alpha$ parameter. Overall, the energy evolution index $\alpha$ for the cases producing significant time delays is found to be in the range $[0.45 - 0.85]$.

        As a consequence, energy-dependent intrinsic time delays obtained from the basic SSC flare model described here evolve with an index $\alpha$  different from the QG model predictions from \citet{ame:1997} and \citet{ell:2000} where $\alpha_{LIV} = 1$. This result illustrates how time-dependent blazar flare scenarios can be used to test these two QG models or any other model predicting a LIV delay with an energy dependency outside the range of values found for the $\alpha$ parameter. Indeed, if a specific QG model predicts a LIV energy-dependent delay with an index $\alpha_{LIV}$ outside this range, the two delays can be discriminated and the QG model can be constrained. Otherwise, the LIV and intrinsic delays are mixed together and remain difficult to disentangle.
        
        \begin{figure}[ht]
                \includegraphics[width=1\linewidth]{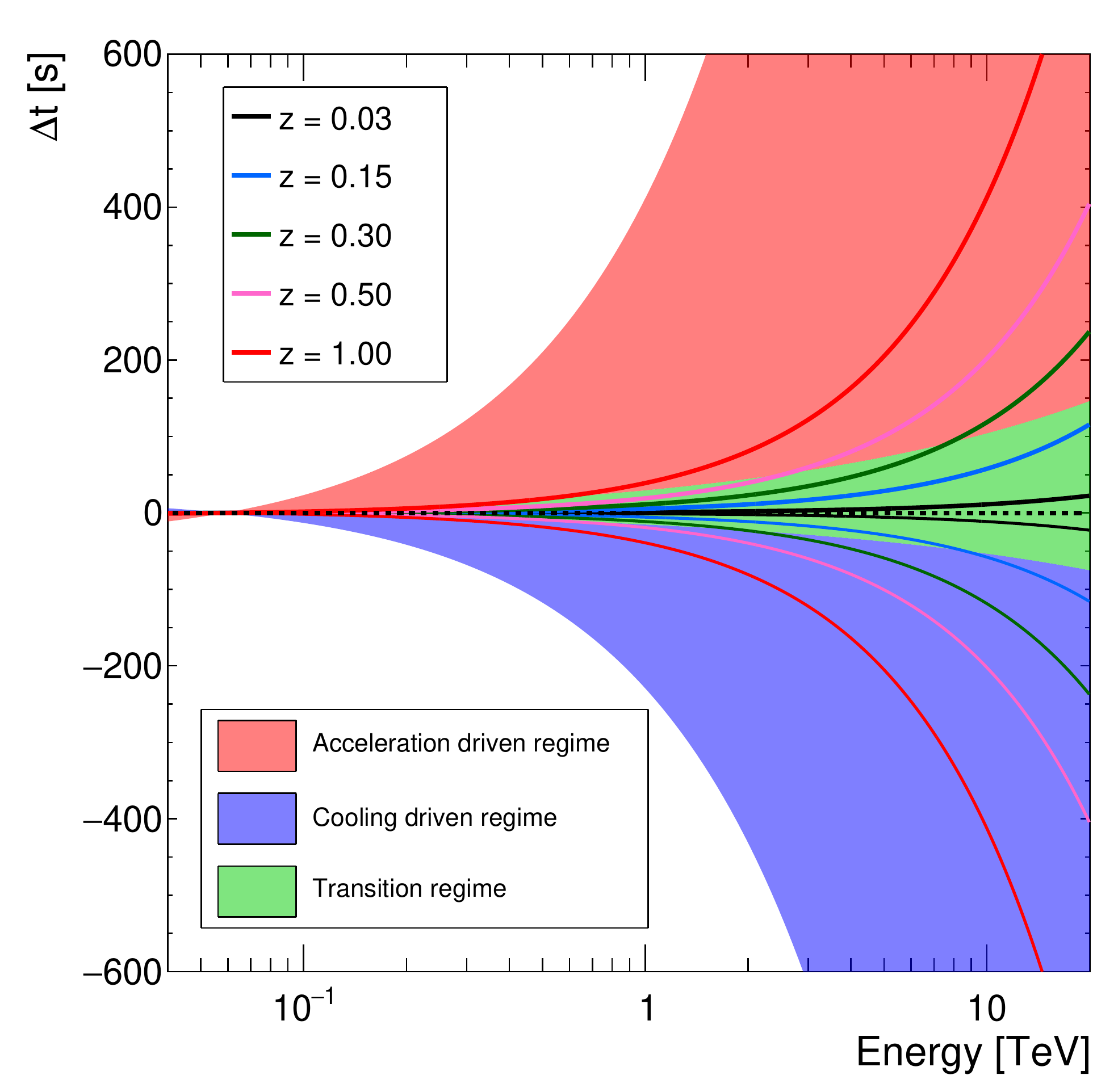}
                \caption{Typical time delays obtained from the SSC flare model (shaded area) and expected from a linear LIV effect with an energy scale at the Planck energy (full lines) with the redshift evolution found in \citet{Jacob2008}. Other formalism such as DSR would lead to different values but of the same order of magnitude. The dashed line corresponds to $\Delta t = 0$. Positive and increasing LIV delays correspond to subluminal LIV effects, negative and decreasing ones to superluminal LIV effects.}
                \label{fig:dt_nrj}
        \end{figure}
        
        To summarize, Figure~\ref{fig:dt_nrj} shows the domain of intrinsic time delays generated in the two regimes and a linear LIV delay at the Planck energy scale ($E_{QG} \sim 10^{19}$~GeV) for subluminal and superluminal effects at different redshifts adopting the redshift evolution from \citet{Jacob2008}. A direct comparison can be done since the redshift does not affect the time-delay evolution with energy but only the observed flux due to the distance and EBL attenuation. Generally speaking, intrinsic time delays appear to be much larger than expected LIV delays in the linear case, except for very large redshifts where observations at VHE remain difficult due to EBL absorption. For higher values of $E_{QG}$, the LIV delay will simply become smaller. In the quadratic case, LIV delays will always be much smaller than intrinsic ones and very difficult to disentangle.

        \renewcommand{\arraystretch}{1.2}
        \begin{table}[ht]
                \begin{center}
                        \begin{tabular}{c | c | c}
                                \hline
                                $B_0$ [mG]      &  $\xi$ [s TeV$^{-\alpha}$]    & $\alpha$                        \\      \hline
                                50                      &       $274 \pm 36$                             & $0.64 \pm 0.1$                \\
                                60                      &       $217 \pm 30$                             & $0.60 \pm 0.1$                \\
                                65                      &       $175 \pm 23$                             & $0.72 \pm 0.1$                \\
                                70                      &       $128 \pm 18$                             & $0.61 \pm 0.1$                \\
                                80                      &       $50 \pm 9$                              & $0.64 \pm 0.2$          \\
                                85                      &       -                                               & -                                       \\
                                90                      &       $-29 \pm 8$                              & $0.57 \pm 0.2$                \\
                                100                     &       $-125 \pm 20$                     & $0.53 \pm 0.1$                \\
                                110                     &       $-181 \pm 23$                     & $0.68 \pm 0.1$        \\ \hline
                        \end{tabular}
                        \vspace{.5cm}
                        \caption{Energy dependent intrinsic time-delay amplitude $\xi$ and power index $\alpha$ for various initial magnetic field strengths in the GeV-TeV energy range. The missing values (shown with a dash) could not be evaluated because there was no significant delay in the considered energy range ($\xi \approx 0$).}
                        \label{tab:alpha_B}
                \end{center}
        \end{table}
        \renewcommand{\arraystretch}{1}

\subsection{Temporal evolution}
        \label{sec:temp_dt}
 A characteristic feature of intrinsic delays is that their magnitude can vary in time during a flare while LIV delays stay constant. This is another observational signature which can provide important information on the origin of the delays. In their study of X-ray variability of blazar flares, \citet{lew:2016} performed a Fourier transform analysis of the delay between two light curves at different energies and obtained the time delay as a function of the Fourier frequency, inversely proportional to the time. Their results show a break in the Fourier transform of the time delay, which occurs at the Fourier frequency corresponding to the time when low-energy photons start to arrive before the high-energy ones. Such a temporal evolution of the delay is a consequence of the mechanisms generating the X-ray flare considered in \citet{lew:2016}. Conversely, the LIV delay is expected to be constant along the flare as it is a cumulative effect over the propagation of all photons from the cosmic source to the Earth. It is entirely determined by the distance and the photon energies emitted by the source.

\begin{figure}[ht]
                \includegraphics[width=.495\linewidth]{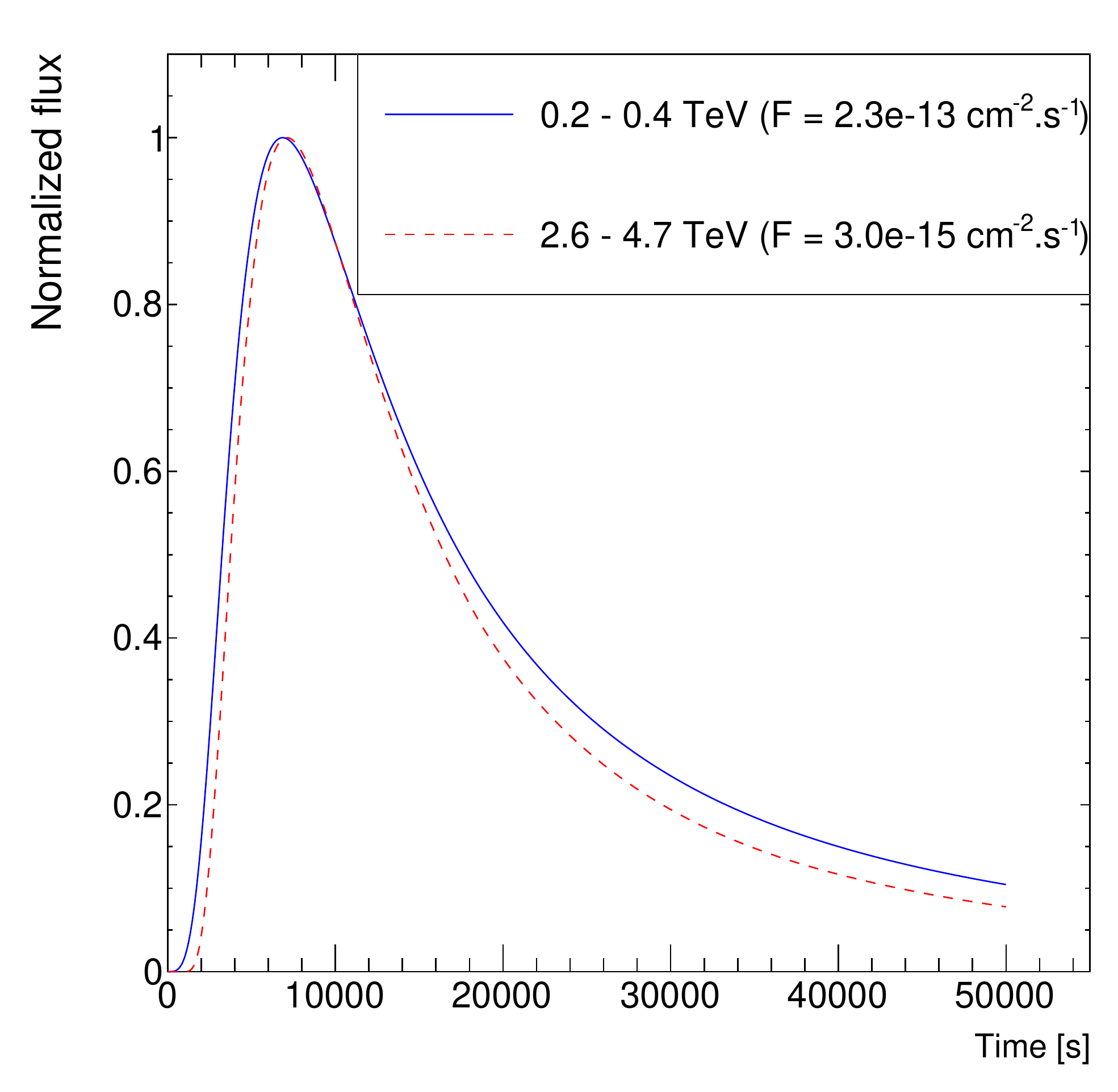} \hfill
                \includegraphics[width=.495\linewidth]{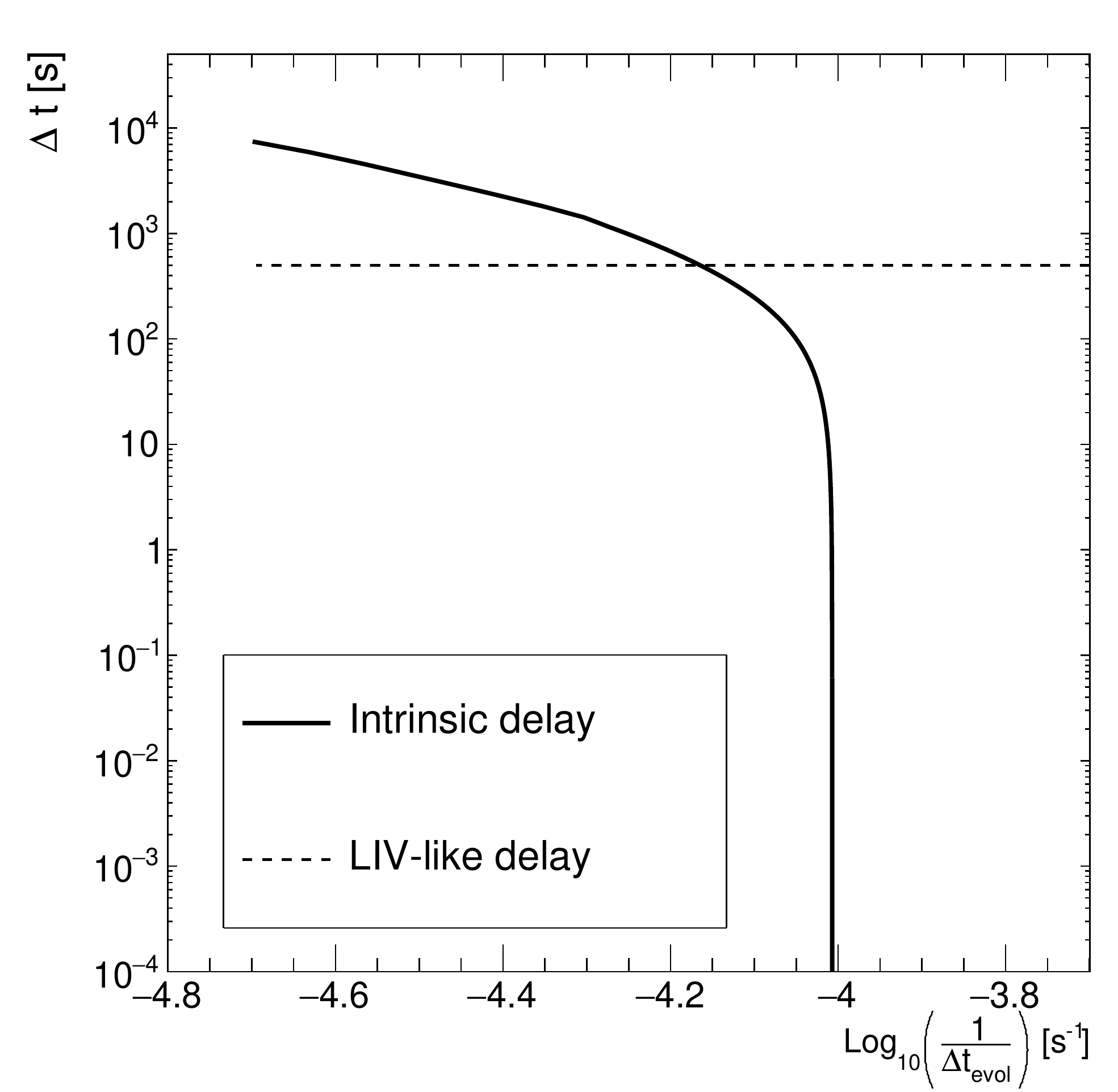}
                \caption{Light curves for a flare with parameters of Table 1 with B = 65 mG for two energy bands (left) and evolution of the simulated time delays between them as a function of time (right). The LIV-like delay overlaid for comparison has been obtained by simulating two asymmetric Gaussian light curves with an injected constant delay of $\Delta t = 500$~s. Only positive delays are shown here in logarithmic scale, which can be directly compared to the \citet{lew:2016} description.}
                \label{fig:dt_four}
        \end{figure}

 To find out the temporal evolution of the delays induced by the SSC model presented here, a simpler method is applied.  The evolution of the delay is evaluated by comparing the time difference $\Delta t_{evol}(t)$ between two light curves reaching the same normalized flux value and by computing this time difference all along the flare. In other words, we noted the normalized fluxes of the two light curves $F_1$ and $F_2$, and considering two times $t$ and $t_2$ such as \mbox{$F_1(t) = F_2(t_2)$}, \mbox{$\Delta t_{\mathrm{evol}}(t) = t-t_2$}. To compare this study with that of \citet{lew:2016}, we have to adopt their opposite sign convention for the time delays, meaning that in this section, a positive delay corresponds to low energies arriving after the high energies. The two light curves chosen for the comparison are integrated over the energy ranges \mbox{$200-400$~GeV} and \mbox{$2.6-4.7$~TeV}. The set of parameters considered here is the one presented in Table~\ref{tab:par_std}, corresponding to an acceleration-driven regime, but results can be obtained in the same  way for the cooling-driven regime. The normalized light curves and the temporal evolution of the delay are shown in Figure~\ref{fig:dt_four}. The result obtained here is similar to that shown in Figure~1 of \citet{lew:2016}. At large times (small Fourier frequencies), the delay is large because the high-energy light curve decays faster than the low-energy one. A break occurs at the specific Fourier frequency corresponding to the time when a given flux is reached earlier for the high-energy light curve than for the low-energy one. At small times (large Fourier frequencies), the delay is negative because the high-energy flare rises after the low-energy one due to the time needed for electrons to be accelerated. Such results are in agreement with the analysis by \citet{lew:2016} and confirm that intrinsic time delays can be significantly variable along the flares and show characteristic time profiles. This variation of the intrinsic time delays explains the result obtained in Section~\ref{sec:tdd}, concerning the CCF which  cannot  correctly reconstruct the injected delay. This is directly due to the fact that the CCF uses the full light curves to evaluate time delays. 

To describe in the same way the situation when only a LIV time delay is present, two light curves were simulated, following the same asymmetric Gaussian shapes as the light curves coming from the SSC model. A constant time delay of $\Delta t = 500$~s is then injected between the two simulated light curves. The resulting temporal evolution of the delay is given in Figure~\ref{fig:dt_four}, which illustrates the constant LIV time delay obtained as a function of the Fourier frequency. Such results clearly show that temporal evolution of time delays is a direct marker for the presence of intrinsic effects in flares observed from blazars. 

\section{Conclusion}
        The blazar flare model considered in this paper describes the evolution of a population of relativistic electrons radiating in a single compact zone through their synchrotron-self-Compton emission. In order to explore basic intrinsic energy-dependent time delays expected in such sources, we focus on a minimal scenario taking into account only the dominant processes needed to generate flares, namely a generic acceleration mechanism and radiative cooling in a slowly varying magnetic field. Under reasonable assumptions, an analytical solution can outline the electron spectrum evolution and applies for instance when synchrotron losses dominate over inverse-Compton losses. Such a scenario clearly emphasizes the likely presence of significant intrinsic time delays. It reveals  the existence of two main time-delay regimes, referred to here as acceleration-driven and cooling-driven regimes, over a large domain of parameters. The system evolves in one of these two regimes depending on the mechanism driving the evolution of the most energetic electrons at the time when the light curves reach their peak and the flare starts to decay. The cooling-driven regime typically corresponds to cases where the decay of the flare is mostly dominated by the radiative cooling effects. Conversely, the acceleration-driven regime corresponds to cases where the acceleration of emitting VHE electrons goes on while the flare starts to decay under the influence of some loss mechanisms other than radiative cooling, such as decrease of the ambient magnetic field or adiabatic losses. The confirmed detection of one of these regimes during blazar monitoring would provide precious information on detailed processes generating flares and significant measurements of time delays would further constrain source parameters \textit{in situ}. 

        However, only upper limits on time delays have been firmly confirmed so far in TeV blazar flares. The detection of possible time delays in a flare of Mrk 501 has remained unique and was observed by only one instrument. Such a situation could be due to the lack of high-quality data on blazar flares. It could also directly put basic flare scenarios into question, which will require further investigation. In the most simple scenarios with fast initial injection or  acceleration of particles immediately followed by flare decay due to radiative losses, a cooling-driven regime could have already been observed during bright flaring events since the cooling time over the VHE domain is  longer than the time resolution reached. Radiative cooling alone should typically induce intrinsic time delays of several minutes. The fact that such lags have not yet been detected suggest that flares could mainly occur in a specific range of parameters, corresponding to an intermediate zone between the two time-delay regimes identified in this paper. As a consequence, there should be a physical link between acceleration and cooling processes with fine-tuning effects during the global evolution of the flare. A possible qualitative scenario would be to consider launching the flare by sudden shock acceleration or magnetic reconnection in the emitting zone, with subsequent mechanisms which induce the flare decay by adiabatic expansion and/or magnetic field decrease, while acceleration processes are still efficient enough at the highest energies. Future observations will be necessary to constrain scenarios and improve time-dependent modeling of blazar flares.
        
        Moreover, the intrinsic delays obtained within the SSC scenario show specific characteristics which could help to constrain QG models or new physics involving time delays. The temporal dependency of the intrinsic delays appears to be different from the one expected by the current description adopted for LIV effects, which may provide a characteristic signature for the presence of intrinsic effects. Indeed, LIV delays are not expected to show any kind of evolution with time since they affect photons in the same way throughout their propagation. On the contrary, intrinsic delays evolve with time due to the different energy-dependent mechanisms involved in the generation of blazar flares. In addition, the energy dependency of intrinsic time delays at GeV-TeV energies was found to present typical power index $\alpha$ in the range $\left[0.45 - 0.85\right]$. This property can be explored in order to test specific QG models which predict energy-dependent LIV delays with an index different from the typical  intrinsic one. Nevertheless, further study of QG models involving LIV effects is necessary to fully exploit the global time-delay information when it becomes available and to distinguish between the various effects. Another important feature to exploit is that LIV delays depend strongly on the propagation distance while intrinsic delays should not.
        
        Briefly, tools and results presented in this paper contribute to the scientific preparation of the new gamma-ray instruments of the coming decade which should provide higher sensitivity and a much larger number of blazar flare detections than current IACTs. Future data will potentially lead to the detection of significant time delays. Flare scenarios should be further developed in order to explain the new observables on intrinsic time delays, and to help disentangle intrinsic and extrinsic effects, opening a way for time-of-flight LIV searches.
        
\bibliography{biblio.bib}
\bibliographystyle{aa}
\hfill

\appendix
\onecolumn
\section{Table for high-energy time delay}
        \renewcommand{\arraystretch}{1.2}
        \begin{table*}[bh]
                \begin{center}
                \begin{tabular}{c | c || c | c || c | c}
                        \hline
                        \multicolumn{2}{c ||}{}                         & \multicolumn{2}{c||}{$B_0 = 65$~mG}  & \multicolumn{2}{c}{$B_0 = 90$~mG}         \\
                                \multicolumn{2}{c ||}{Parameter}                & $\xi$ [s TeV$^{-\alpha}$]       & $\alpha$                              &  $\xi$ [s TeV$^{-\alpha}$]      & $\alpha$                                              \\ \hline
                                 $m_\mathrm{b}$         & \bfseries{1.0}                         & $\bold{175 \pm 23}$           & $\bold{0.72 \pm 0.1}$   & $\bold{-29 \pm 8}$                    & $\bold{0.57 \pm 0.2}$                 \\
                                                & 1.25                                          & $300 \pm 39$                            & $0.61 \pm 0.1$                        & $154 \pm 21$                            & $0.65 \pm 0.1$                                        \\
                                                & 1.5                                           & $369 \pm 59$                            & $0.49 \pm 0.2$                        & $226 \pm 29$                            & $0.63 \pm 0.1$                                        \\
                                                & 2.0                                            & $391 \pm 62$                          & $0.52 \pm 0.2$                    & $365 \pm 54$                          & $0.53 \pm 0.2$                                    \\ \hline
                                $\delta$        & 20                                            & $413 \pm 90$                            & $0.68 \pm 0.5$                        & $-97 \pm 15$                            & $0.63 \pm 0.3$                                        \\
                                                & 30                                            & $227 \pm 32$                            & $0.59 \pm 0.2$                        & $-46 \pm 10$                            & $0.62 \pm 0.3$                                        \\
                                                & \bfseries{40}                         & $\bold{175 \pm 23}$             & $\bold{0.72 \pm 0.1}$ & $\bold{-29 \pm 8}$                     & $\bold{0.57 \pm 0.2}$                 \\
                                                &50                                                     & $96 \pm 14$                             & $0.47 \pm 0.1$                        & $-26 \pm 10$                            & $0.49 \pm 0.3$                                        \\ \hline
                                $A_0$   &$4.0\times10^{-5}$                     & $264 \pm 36$                            & $0.61 \pm 0.2$                        & $69 \pm 12$                             & $0.60 \pm 0.2$                                        \\
                        $\left[s^{-1}\right]$   &$\bold{4.5\times10^{-5}}$         & $\bold{175 \pm 23}$           & $\bold{0.72 \pm 0.1}$ & $\bold{-29 \pm 8}$                 & $\bold{0.57 \pm 0.2}$                 \\
                                                &$5.0\times10^{-5}$                     & $81 \pm 14$                             & $0.50 \pm 0.1$                        & $-125 \pm 18$                           & $0.63 \pm 0.1$                                        \\
                                                &$5.5\times10^{-5}$                     & -                                                       & -                                             & $-202 \pm 28$                           & $0.57 \pm 0.1$                                        \\
                                                &$6.0\times10^{-5}$                      & $-72 \pm 11$                          & $0.58 \pm 0.1$                        & $-225 \pm 27$                           & $0.80 \pm 0.1$                                        \\ \hline          
                                $m_a$   & 4.7                                           & $51 \pm 9$                                      & $0.62 \pm 0.1$                        & $-205 \pm 28$                           & $0.59 \pm 0.1$                                        \\
                                                & 5.0                                           & $94 \pm 13$                             & $0.65 \pm 0.1$                        & $-106 \pm 15$                           & $0.66 \pm 0.1$                                        \\
                                                & 5.3                                           & $158 \pm 24$                            & $0.48 \pm 0.2$                        & $-75 \pm 12$                            & $0.67 \pm 0.2$                                        \\
                                                & \bfseries{5.6}                        & $\bold{175 \pm 23}$             & $\bold{0.72 \pm 0.1}$ & $\bold{-29 \pm 8}$                     & $\bold{0.57 \pm 0.2}$                 \\
                                                & 5.9                                           & $190 \pm 26$                            & $0.61 \pm 0.2$                        & -                                                       & -                                                             \\ \hline
                                $n$             & 2.2                                           & -                                                       & -                                             & $52 \pm 11$                             & $0.43 \pm 0.1$                                        \\
                                                & 2.3                                           & $173 \pm 22$                            & $0.80 \pm 0.1$                        & -                                                       & -                                                             \\
                                                & \bfseries{2.4}                        & $\bold{175 \pm 23}$             & $\bold{0.72 \pm 0.1}$ & $\bold{-29 \pm 8}$                     & $\bold{0.57 \pm 0.2}$                 \\
                                                & 2.5                                           & $277 \pm 153$                           & $0.26 \pm 0.3$                        & $-72 \pm 20$                            & $0.44 \pm 0.2$                                        \\
                                                & 2.6                                           & $113 \pm 26$                            & $0.66 \pm 0.6$                        & $-104 \pm 37$                           & $0.81 \pm 0.7$                                        \\
                                                & 2.7                                           & -                                                       &       -                                       & -                                                       &       -                                                       \\
                                                & 2.8                                           & -                                                       &       -                                       & -                                                       &       -                                                       \\ \hline
                $\gamma_{c,0}$  & $2\times10^{4}$                       & $401 \pm 51$                            & $0.64 \pm 0.2$                        & $241 \pm 31$                            & $0.66 \pm 0.2$                                        \\
                                                & $3\times10^{4}$                       & $265 \pm 37$                            & $0.55 \pm 0.2$                        & $57 \pm 9$                              & $0.85 \pm 0.2$                                        \\
                                                & $\bold{4\times10^{4}}$         & $\bold{175 \pm 23}$           & $\bold{0.72 \pm 0.1}$ & $\bold{-29 \pm 8}$                 & $\bold{0.57 \pm 0.2}$                 \\
                                                & $5\times10^{4}$                       & $84 \pm 13$                             & $0.57 \pm 0.1$                        & $-82 \pm 12$                            & $0.78 \pm 0.2$                                        \\
                                                & $6\times10^{4}$                       & $61 \pm 14$                             & $0.42 \pm 0.1$                        & $-131 \pm 19$                           & $0.61 \pm 0.1$                                        \\
                                                & $7\times10^{4}$                       & -                                                       & -                                             & $-151 \pm 21$                           & $0.63 \pm 0.1$                                        \\
                                                & $8\times10^{4}$                       & -                                                       & -                                     & $-205 \pm 30$                           & $0.54 \pm 0.1$                                        \\ \hline
                                \end{tabular}
                                \vspace{.5cm}
                                \caption[Energy dependent intrinsic time delay amplitude $\xi$ and power index $\alpha$ for all investigated parameter values for the two initial magnetic field strength values.]{Energy-dependent time-delay index for all investigated parameters for the two initial magnetic-field-strength values taken as benchmark values. Bold lines correspond to these benchmark cases, given by the parameters in Table~\ref{tab:par_std}. The missing values (shown with a dash) could not be evaluated because there was no significant delay in the considered energy range or because the maximum energy emitted was below 250~GeV.}
                \label{tab:alpha_all}
                \end{center}
        \end{table*}
\renewcommand{\arraystretch}{1}

\end{document}